\documentclass[12pt]{article}
\usepackage{putex}

\makeatletter
\DeclareRobustCommand*{\bfseries}{%
  \not@math@alphabet\bfseries\mathbf
  \fontseries\bfdefault\selectfont
  \boldmath
}
\makeatother

\input{glyphtounicode}
\newcommand{\Gammafct}{\operatorname{\Gamma}}
\newcommand{\zetafct}{\operatorname{\zeta}}
\def\extremum{\mathop{\rm extremum}\displaylimits}

\begin{document}

\preprint{PUPT-2503}

\title{$p$-adic AdS/CFT}
\authors{Steven S. Gubser, Johannes Knaute\footnote{Also at the Institut f\"ur Theoretische Physik, TU Dresden, 01062 Dresden, Germany}, Sarthak Parikh,\\[5pt] Andreas Samberg\footnote{Also at the Institut f\"ur Theoretische Physik, Ruprecht-Karls-Universit\"at Heidelberg, Philosophenweg 16, 69120 Heidelberg, Germany, and at the ExtreMe Matter Institute EMMI, GSI Helmholtzzentrum f\"ur Schwerionenforschung, Planckstra{\ss}e~1, 64291~Darmstadt, Germany}, and Przemek Witaszczyk\footnote{Also at the Institute of Physics, Jagiellonian University, S. Lojasiewicza 11, 30-348 Krakow, Poland}}
\institution{PU}{Joseph Henry Laboratories, Princeton University, Princeton, NJ 08544, USA}

\abstract{We construct a $p$-adic analog to AdS/CFT, where an unramified extension of the $p$-adic numbers replaces Euclidean space as the boundary and a version of the Bruhat--Tits tree replaces the bulk.  Correlation functions are computed in the simple case of a single massive scalar in the bulk, with results that are strikingly similar to ordinary holographic correlation functions when expressed in terms of local zeta functions.  We give some brief discussion of the geometry of $p$-adic chordal distance and of Wilson loops.  Our presentation includes an introduction to $p$-adic numbers.}

\date{May 2016}
\maketitle

\tableofcontents

\section{Introduction}
\label{INTRODUCTION}

In the anti-de Sitter (AdS) / conformal field theory (CFT) correspondence \cite{Maldacena:1997re,Gubser:1998bc,Witten:1998qj} (for a review, see \cite{Aharony:1999ti}), classical supergravity in an asymptotically ${\rm AdS}_{n+1}$ geometry is dual to a large $N$ gauge theory on the boundary of ${\rm AdS}_{n+1}$, which is $\mathbb{R}^n$, or more precisely its projective completion $S^n$.  From the perspectives both of tensor networks in AdS/CFT \cite{Swingle:2009bg,Swingle:2012wq,Qi:2013caa,Pastawski:2015qua} and of segmented strings \cite{Vegh:2015ska,Callebaut:2015fsa,Vegh:2015yua,Gubser:2016wno,Gubser:2016zyw,Vegh:2016hwq,Vegh:2016fcm}, it is desirable to consider analogs of AdS/CFT based on discrete rather than continuous bulk geometries.  A similar development played out in the context of $p$-adic string theory \cite{Freund:1987kt,Freund:1987ck}, where the worldsheet of the open string is replaced \cite{Zabrodin:1988ep} by the so-called Bruhat--Tits tree \cite{Bruhat:1972} (known more commonly to physicists as the Bethe lattice with coordination number $p+1$), whose boundary is not $S^1$ but instead the projective space $\mathbb{P}^1(\mathbb{Q}_p)$, where $\mathbb{Q}_p$ is the field of $p$-adic numbers.  Relevance of $p$-adic numbers and the Bruhat--Tits tree to dS/CFT was pointed out in \cite{Harlow:2011az} in the context of a statistical mechanical model capturing aspects of eternal inflation.  Here we will develop what appears to us a closer connection between the Bruhat--Tits tree for $\mathbb{Q}_p$ and the Euclidean ${\rm AdS}_2/{\rm CFT}_1$ duality, based on realizing the Bruhat--Tits tree as a group quotient analogous to ${\rm SL}(2,\mathbb{R})/{\rm U}(1)$.  In retrospect we can recognize the derivation in \cite{Zabrodin:1988ep} of an effective non-local action on the boundary from a classical theory on the Bruhat--Tits tree as in the spirit of ${\rm AdS}_2/{\rm CFT}_1$.

The bulk of this paper is devoted to a detailed account of how to formulate AdS/CFT when the boundary is not $\mathbb{R}^n$ but instead an $n$-dimensional vector space $\mathbb{Q}_p^n$.  More precisely, we take the boundary to be the unramified extension of $\mathbb{Q}_p$ of degree $n$, and the bulk is a modification of the Bruhat--Tits tree for $\mathbb{Q}_p$ such that each vertex has $p^n+1$ nearest neighbors.\footnote{Quadratic field extensions appeared already in \cite{Freund:1987kt} in connection with closed strings, which is to say two-dimensional conformal field theory.  Ramified extensions have been considered in the context of the $p$-adic string in \cite{Ghoshal:2006zh}, following work of \cite{Gerasimov:2000zp}, and other field extensions of the $p$-adics have also been considered for some time in the program of Gervais \cite{Gervais:1987we} to generalize $p$-adic string amplitudes.}  For simplicity, we limit the discussion to the simplest possible action on the modified Bruhat--Tits tree, namely a nearest neighbor action for a single real scalar with a mass and possibly some cubic or quartic self-interactions.  Green's functions can be expressed in terms of bulk-to-boundary and bulk-to-bulk propagators, and a number of formal similarities to ordinary AdS/CFT can be noted.  Final expressions for two- and three-point correlators are mostly, but not entirely, amenable to being assembled into adelic products.  The main feature of $p$-adic four-point amplitudes is a remarkably simple closed form expression for the $p$-adic amplitudes as compared to their real counterparts.

The organization of the rest of this paper is as follows.  In section~\ref{PADIC} we summarize the aspects of $p$-adic numbers that we will need.  Most of our account is standard, and more thorough treatments can be found, for example, in \cite{brekke1988non,gouvea1997p}.  In section~\ref{PROPAGATORS}, we introduce the classical scalar dynamics that we are interested in on the Bruhat--Tits tree, and we explain how to compute bulk-to-bulk and bulk-to-boundary propagators.  We then pass in section~\ref{CORRELATORS} to the computation of holographic $m$-point amplitudes for $m=2$, $3$, and $4$.  We conclude with a discussion in section~\ref{DISCUSSION} touching on the relation of $p$-adic correlators to their real counterparts, the geometry of chordal distance on $p$-adic ${\rm AdS}_{n+1}$, and future directions.

\section{$p$-adic numbers}
\label{PADIC}

Let $p$ be a prime number.  A non-zero $p$-adic number is a series
 \eqn{padicDecomp}{
  x = p^{v_p(x)} \sum_{m=0}^\infty a_m p^m \,,\quad a_0\ne 0\,,
 }
where the $a_m\in\{0,1,\dots,p-1\}$ are digits and $v_p(x)\in\mathbb{Z}$ is called the $p$-adic valuation of $x$.  The $p$-adic absolute value or norm of $x$ is then defined as
 \eqn{padicNorm}{
  |x|_p = p^{-v_p(x)} \,.
 }
The set of all numbers $x$ of the form \eno{padicDecomp}, together with $0$, is denoted $\mathbb{Q}_p$.  We define $|0|_p = 0$, and correspondingly $v_p(0) = \infty$.  The sum in \eno{padicDecomp} is convergent with respect to the norm $|\cdot|_p$.  Intuitively, the $p$-adic norm is based on regarding $p$ as small but non-zero, while integers prime to $p$ are all the same size.  To define addition and multiplication on $\mathbb{Q}_p$, we can define them in the standard manner (as rationals) for series that terminate, and then extend the definition to all of $\mathbb{Q}_p$ by insisting that addition and multiplication should be continuous maps with respect to $|\cdot|_p$.

It is easy to show that every element of $\mathbb{Q}_p$ has an additive inverse, so that subtraction is defined.  We first note that
 \eqn{MinusOne}{
  -1 &= {p-1 \over 1-p} = (p-1) \left( 1 + p + p^2 + p^3 + \ldots \right)  \cr
    &= (p-1) + (p-1) p + (p-1) p^2 + (p-1) p^3 + \ldots \,,
 }
and the expression in the second line takes the form \eno{padicDecomp} with $v_p(-1)=0$ and all $a_m = p-1$.  The second equality uses the standard geometric series expansion for $1/(1-p)$, and it is justified because $p$ is smaller than $1$ in the $p$-adic norm $|\cdot|_p$.  Multiplication of any element of $\mathbb{Q}_p$ by $-1$ gives its unique additive inverse.  It can be shown that unique multiplicative inverses also exist for non-zero $p$-adic numbers.  In short, $\mathbb{Q}_p$ is a field.  It can be constructed rigorously as the completion of the rationals $\mathbb{Q}$ with respect to $|\cdot|_p$.

The set $\mathbb{Z}_p$ of $p$-adic integers is defined to be all elements of $\mathbb{Q}_p$ with $|x|_p \leq 1$, and it is a ring: That is, we can add, subtract, and multiply $p$-adic integers to get new $p$-adic integers, but multiplicative inverses do not always exist in $\mathbb{Z}_p$.  The set of units $\mathbb{U}_p$ in $\mathbb{Z}_p$ is all elements of $\mathbb{Z}_p$ with $|x|_p = 1$, and precisely these $p$-adic integers {\it do} have multiplicative inverses in $\mathbb{Z}_p$.  We may re-express \eno{padicDecomp} as
 \eqn{padicDecompAgain}{
  x = p^{v_p(x)} \hat{x} \,,
 }
where $\hat{x}$ is a unit, uniquely determined by non-zero $x$.  Intuitively, we think of $\mathbb{Z}_p$ as the unit ball in $\mathbb{Q}_p$, while $\mathbb{U}_p$ is the unit sphere.  Because the decomposition \eno{padicDecompAgain} is unique, we may express the non-zero $p$-adic numbers as
 \eqn{Qexpress}{
  \mathbb{Q}_p^\times = \bigsqcup_{m \in \mathbb{Z}} p^m \mathbb{U}_p \,,
 }
where $\sqcup$ indicates a disjoint union\footnote{We will have enough occasion in this paper to use both ordinary unions, represented by $\cup$, and disjoint unions, represented by $\sqcup$, that we emphasize the distinction.  By definition,
 \eqn{TwoUnions}{
 \bigcup_{\alpha \in S} A_\alpha \equiv \left\{ x : x\in A_\alpha \hbox{ and } \alpha \in S \right\} \qquad\hbox{while}\qquad
 \bigsqcup_{\alpha \in S} A_\alpha \equiv \left\{  (x, \alpha) : x \in A_\alpha \hbox{ and } \alpha \in S \right\}\,.
 } 
Note that the ordinary union and disjoint union of sets can be naturally identified if and only if all the sets $A_\alpha$ are disjoint.  Sometimes, as in \eno{Qexpress}, we use $\sqcup$ to indicate a union of obviously disjoint sets; elsewhere, as in \eno{DisjointUnion}, we use $\sqcup$ on overlapping sets when the multiplicity of elements in the final union matters.  We reserve $\cup$ for use in situations where we want the ordinary union of sets which may overlap.} and $\mathbb{Q}_p^\times$ is the multiplicative group of units in $\mathbb{Q}_p$, namely $\mathbb{Q}_p \;\backslash\; \{0\}$.

The $p$-adic norm $|\cdot|_p$ is ultrametric, meaning that for $x,y \in \mathbb{Q}_p$, we have $|x+y|_p \leq \sup\{|x|_p,|y|_p\}$.  An important corollary of this property is the so-called tall isosceles property, which says that if $x+y+z=0$ in $\mathbb{Q}_p$, then after relabeling $x$, $y$, and $z$ if necessary, we always have $|x|_p = |y|_p \geq |z|_p$.

\subsection{The Bruhat--Tits tree}
\label{BTTREE}

\begin{figure}
  \centerline{\includegraphics[width=4in]{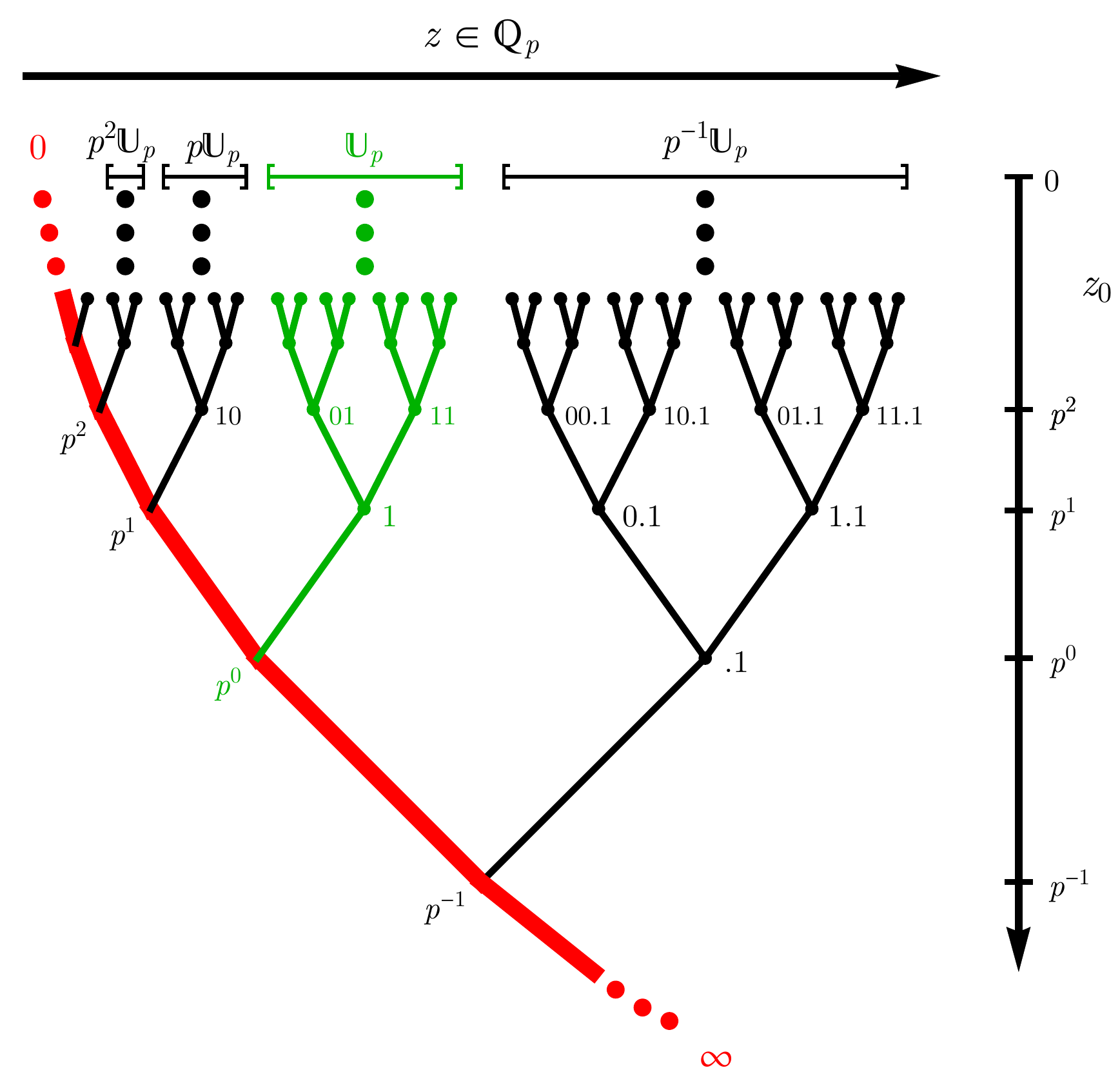}}
  \caption{The Bruhat--Tits tree for $\mathbb{Q}_p$ with $p=2$, with a coordinate system $(z_0, z)$ shown which provides a useful parametrization of the bulk-to-boundary propagator.}\label{BruhatTits}
 \end{figure}

There is a well-known relation between the $p$-adic numbers and trees.  Consider first an element of $\mathbb{U}_p$.  The first (rightmost) digit is non-zero, so there are $p-1$ choices for it.  Once that choice is made, there are $p$ choices for the next digit, and the next, and so forth.  A convenient graphical way to depict the relation \eno{Qexpress} is to show the sets $p^m \mathbb{U}_p$ as bushes rooted in a trunk, with each root corresponding to some fixed power $p^m$.  Each non-zero $p$-adic number $x$ is the terminus of a unique upward path through one of the bushes, and the magnitude $|x|_p$ corresponds to the bush in which the path lies.  It is natural to go further and include points $0$ and $\infty$ as the terminal points on each end of the trunk.  Altogether, the resulting graph is the Bruhat--Tits tree, $T_p$, which is a regular tree with coordination number $p+1$.  See figure~\ref{BruhatTits}.

The boundary of the Bruhat--Tits tree is $\mathbb{Q}_p \sqcup \{\infty\}$, which more properly is the projective space $\mathbb{P}^1(\mathbb{Q}_p)$.  We can realize $\mathbb{P}^1(\mathbb{Q}_p)$ as all pairs $(x,y) \in \mathbb{Q}_p^2 \;\backslash\; \{(0,0)\}$ modulo the relation $(x,y) \sim (\lambda x,\lambda y)$ for $\lambda \in \mathbb{Q}_p^\times$.  There is a natural action of ${\rm PGL}(2,\mathbb{Q}_p)$ on $\mathbb{P}^1(\mathbb{Q}_p)$, most simply realized as linear fractional transformations
 \eqn{padicLFTs}{
  x \to {\alpha x + \beta \over \gamma x + \delta} \,,
 }
where $\alpha$, $\beta$, $\gamma$, and $\delta$ are all elements in $\mathbb{Q}_p$, satisfying $\alpha\delta - \beta\gamma \neq 0$.  A slightly more careful definition is first to introduce ${\rm GL}(2,\mathbb{Q}_p)$, which is all matrices $M = \scriptsize{\begin{pmatrix} \alpha & \beta \\ \gamma & \delta \end{pmatrix}}$ with entries in $\mathbb{Q}_p$ and with $\alpha\delta - \beta\gamma \neq 0$; and then to identify elements $M$ and $\tilde{M}$ of ${\rm GL}(2,\mathbb{Q}_p)$ whenever $M = \lambda\tilde{M}$ for some $\lambda \in \mathbb{Q}_p^\times$.  In short, ${\rm PGL}(2,\mathbb{Q}_p) = {\rm GL}(2,\mathbb{Q}_p) / \mathbb{Q}_p^\times$.  Likewise we can define ${\rm PGL}(2,\mathbb{Z}_p) = {\rm GL}(2,\mathbb{Z}_p) / \mathbb{U}_p$ (recalling that $\mathbb{U}_p$ is the group of units in $\mathbb{Z}_p$) and a remarkable fact is that the Bruhat--Tits tree, $T_p$, is naturally identified as ${\rm PGL}(2,\mathbb{Q}_p) / {\rm PGL}(2,\mathbb{Z}_p)$.

By intent, the Bruhat--Tits tree is a discrete analog of a Riemannian symmetric space.  The analogy can be summarized in tabular form as follows:
 \eqn{BTanalogy}{\seqalign{\span\TC\quad & \quad\span\TC\quad & \quad\span\TC\quad & \quad\span\TC}{
  \hbox{symmetry group} & \hbox{maximal compact subgroup} & \hbox{quotient space} & \hbox{boundary}
    \cr\noalign{\vskip1\jot}\hline\noalign{\vskip1\jot}
  {\rm PGL}(2,\mathbb{Q}_p) & {\rm PGL}(2,\mathbb{Z}_p) & T_p & \mathbb{P}^1(\mathbb{Q}_p)
    \cr
  {\rm SL}(2,\mathbb{R}) & {\rm SO}(2,\mathbb{R}) & \mathbb{D} & S^1
 }}
where $\mathbb{D}$ is the Poincar\'e disk.  We should be encouraged by this table to think that some $p$-adic version of the $\hbox{AdS}_2/\hbox{CFT}_1$ correspondence can be formulated.  Indeed, the derivation in \cite{Zabrodin:1988ep} of a non-local effective action on the boundary can be recognized in retrospect as a holographic calculation of two-point functions for operators related to a massless scalar in the bulk.  In section~\ref{CORRELATORS} we will develop $p$-adic $\hbox{AdS}_2/\hbox{CFT}_1$ by computing connected two-, three-, and four-point correlators from a classical discrete action defined on the Bruhat--Tits tree.  For example, two-point correlators take the form $\langle {\cal O}(x) {\cal O}(0) \rangle = F^{(2)}_p/|x|_p^{2\Delta}$ where $\Delta$ is the dimension of ${\cal O}$ and $F^{(2)}_p$ is a normalization factor.

We will find it natural to refer to a particular depth coordinate $z_0$ on the Bruhat--Tits tree, where $z_0 = p^\omega$ and $\omega \in \mathbb{Z}$.  Along the trunk of the tree, $z_0$ is $p^m$ at the point where the bush terminating in $p^m \mathbb{U}_p$ is rooted.  At a node of the tree not on the trunk, we have chosen some finite number of $p$-adic digits, and $z_0$ is the first power of $p$ corresponding to a digit we have {\it not} chosen.  For instance, at the point $10.1$ on the $2$-adic tree shown in figure~\ref{BruhatTits}, $z_0 = p^2$ because in writing $10.1$ we have specified the $1/p$-place digit, the ones-place digit, and the $p$-place digit, but not the $p^2$-place digit.  In short, $z_0$ is $p$-adic accuracy.  More formally, any point on the tree can be considered an equivalence class of $p$-adic numbers, where the equivalence relation is equality up to $O(z_0)$ corrections.  The equivalence classes all take the form $z + z_0 \mathbb{Z}_p$, where $z \in \mathbb{Q}_p$. To say it another way: if $z$ is any $p$-adic number, then the point $(z_0,z)$ on the tree is the point whose digits up to $p^{v_p(z_0)}$ match the corresponding digits of $z$.  This even works for points along the trunk if we think of all their chosen digits as $0$.\footnote{To bring this discussion closer to the standard mathematical description of the Bruhat--Tits tree (see e.g.~\cite{brekke1988non}), we can rephrase our definition of a point on $T_p$ so that each point is an equivalence class in $\mathbb{Q}_p \times \mathbb{Q}_p$, where elements $(z_0,z)$ and $(z_0',z')$ in $\mathbb{Q}_p \times \mathbb{Q}_p$ are identified if $|z_0| = |z_0'|$ and $z \in z' + z_0' \mathbb{Z}_p$.}

\subsection{Field extensions}

\begin{figure}
  \begin{picture}(500,200)
   \put(120,-30){\includegraphics[width=3in]{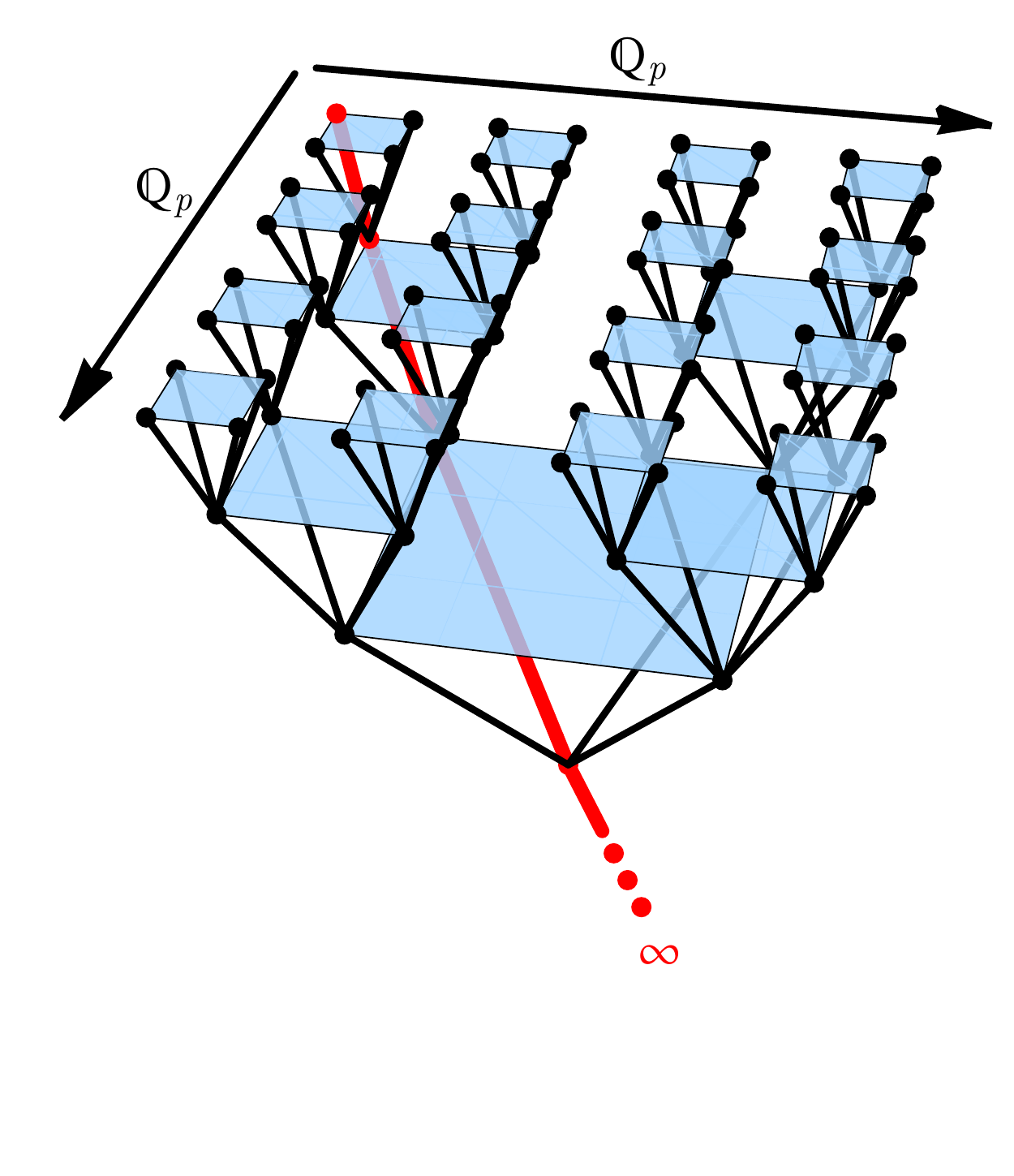}}
  \end{picture}
  \caption{A variant of the Bruhat--Tits tree for the unramified extension $\mathbb{Q}_{p^2}$ with $p=2$.}\label{BruhatTits2}
 \end{figure}

We would like to inquire whether there is a natural generalization of the Bruhat--Tits tree for $\mathbb{Q}_p$ which allows us to formulate a $p$-adic version of $\hbox{AdS}_{n+1}/\hbox{CFT}_n$.  For $n=2$, our goal is captured pictorially in figure~\ref{BruhatTits2}.
To discuss $n$-dimensional $p$-adic field theories, we need to define some natural norm $|x|$ for vectors $x$ in a vector space $\mathbb{Q}_p^n$.  This norm will enter into correlators: For example, we expect $\langle {\cal O}(x) {\cal O}(0) \rangle = F^{(2)}_{p^n} / |x|^{2\Delta}$ for some normalization constant $F^{(2)}_{p^n}$.

There are in fact natural norms on the vector space $\mathbb{Q}_p^n$.  A surprise, however, is that they are formulated rather differently from the usual ${\rm O}(n)$ symmetric norm on $\mathbb{R}^n$, and they are non-unique.  To construct these norms, we start with a field extension $K$ of $\mathbb{Q}_p$ of degree $n$.\footnote{A simple example of a field extension of degree $2$ is $\mathbb{Q}(\sqrt{2})$, which is the smallest field including both $\mathbb{Q}$ and $\sqrt{2}$.  Any element may be expressed as $z = x + y \sqrt{2}$, and the natural field norm is $N(z) = x^2 - 2y^2$.  In this case, $N(z) = z\bar{z}$ where $\bar{z} = x - y \sqrt{2}$ can be characterized as the conjugate of $z$.  In general, if a larger field $K$ contains a smaller field $L$ and can be represented as an $n$-dimensional vector space over $L$, then $K$ is a field extension of $L$ of degree $n$.  Given an element $a \in K$, the map $v \to av$ for any other element $v \in K$ amounts to a linear map on $L^n$, and so we can calculate its trace $\Tr_{K:L}(a)$ and its determinant $N_{K:L}(a)$.  The determinant of this map is the field norm $N(a)$, or in more precise notation $N_{K:L}(a)$ where $K:L$ specifies the field extension under consideration.  $N_{K:L}$ is a homogeneous map from $K$ to $L$ of degree $n$, in the sense that $N_{K:L}(\lambda a) = \lambda^n N_{K:L}(a)$ when $\lambda \in L$.  Note that $N_{K:L}(1) = 1$ because multiplying an element of $L$ by $1$ is represented as multiplying the associated vector in $K^n$ by the $n \times n$ identity matrix.}  Using the associated field norm $N(x)$ for $x \in K$, we can define a norm 
 \eqn{FieldExtensionNorm}{
  |x|_K = |N(x)|_p^{1/n} \,.
 }
This is the unique norm on $K$ satisfying $|x|_K = |x|_p$ for any $x \in \mathbb{Q}_p$ and $|xy|_K = |x|_K |y|_K$ for any $x,y \in K$.  Since $K = \mathbb{Q}_p^n$ as a vector space, \eno{FieldExtensionNorm} defines a natural norm on $\mathbb{Q}_p^n$, and it turns out to be an ultrametric norm; so in particular it satisfies the tall isosceles property. Different field extensions of the same degree define inequivalent norms on $\mathbb{Q}_p^n$.  We will be most interested in the unramified extension of degree $n$, which we will denote by $\mathbb{Q}_{p^n}$.  By definition, it is the field extension (which turns out to be unique) such that $|x|_K$ as defined in \eno{FieldExtensionNorm} is always an integer power of $p$ for non-zero $x$.  Other field extensions of $\mathbb{Q}_p$ can be labeled (though not always uniquely) by the smallest integer divisor $e$ of $n$ such that $|x|_K^e$ is an integer power of $p$ for all non-zero $x$.  One refers to $e$ as the ramification index, and $e=1$ corresponds to the unramified case.

Given a field extension $K$ of $\mathbb{Q}_p$ (not necessarily unramified), we can introduce analogs of $\mathbb{Z}_p$ and $\mathbb{U}_p$, namely
 \eqn{OKdef}{
  \mathbb{Z}_K \equiv \{ x \in K\colon |x|_K \leq 1 \} \qquad\hbox{and}\qquad 
  \mathbb{U}_K \equiv \{ x \in K\colon |x|_K = 1 \} \,.
 }
Like the $p$-adic integers, $\mathbb{Z}_K$ is a ring but not a field.  If we define $\mathfrak{p}_K \equiv \{ x \in K\colon |x|_K < 1 \}$, then it can be shown that $\mathfrak{p}_K$ is a maximal ideal in $\mathbb{Z}_K$, so that $\mathbb{Z}_K/\mathfrak{p}_K$ is a field, called the residue field.  It is in fact the finite field $\mathbb{F}_{p^f}$ with $p^f$ elements, where $f = n/e$.  It can also be shown that one can find an element $\pi \in K$ with $|\pi|_K = p^{-1/e}$, and that once such an element (called a ``uniformizer'') is chosen, the polar decomposition summarized in \eno{padicDecomp} and \eno{padicDecompAgain} can be generalized to a unique representation of any non-zero element of $K$:
 \eqn{qadicDecomp}{
  x = \pi^{v_K(x)} \sum_{m=0}^\infty a_m \pi^m \,,\quad a_0\ne 0\,,
 }
where $v_K(x) \in \mathbb{Z}$ is the valuation of $x$ in $K$, and the $a_k$ are elements of the residue field, with $a_0 \neq 0$.  From the perspective of a tree representation, we see from \eno{qadicDecomp} that we can represent the sets $\pi^{m} \mathbb{U}_K$ as bushes rooted in a trunk, with each root marked by a power $\pi^m$ of the uniformizer.  Starting on the trunk, the first step up into a chosen bush amounts to choosing $a_0 \neq 0$, and subsequent steps amount to choosing successive ``digits'' $a_k$ in the residue field.  We see that the tree has uniform coordination number $p^f+1$.

Two examples may help make the discussion of the previous paragraph clearer.  First, the totally ramified extension of $\mathbb{Q}_p$ of degree $n$ comes from extending $\mathbb{Q}_p$ by $p^{1/n}$.  Then $e=n$ and $f=1$, and the uniformizer can be chosen as $p^{1/n}$ itself.  The tree is identical to the original Bruhat--Tits tree, only the natural depth coordinate $z_0$ now takes values $p^{m/n}$ where $m \in \mathbb{Z}$.  As argued in \cite{Ghoshal:2006zh}, it can be thought of as a refinement of the Bruhat--Tits tree for $\mathbb{Q}_p$ itself.  Second, the unique unramified extension $\mathbb{Q}_{p^n}$ of $\mathbb{Q}_p$ of degree $n$ can be obtained by adjoining to $\mathbb{Q}_p$ a primitive $(p^n-1)$-th root of unity.  This non-trivial assertion is demonstrated, for example, on pp.~167ff of \cite{gouvea1997p}.  The obvious choice of uniformizer is $p$, and the natural tree structure associated with $\mathbb{Q}_{p^n}$ is shown in figure~\ref{BruhatTits2}.  The depth coordinate $z_0$ takes values $p^\omega$ with $\omega \in \mathbb{Z}$.  The boundary is $\mathbb{Q}_{p^n} \sqcup \{\infty\}$, which more properly is $\mathbb{P}^1(\mathbb{Q}_{p^n})$.  We will denote this modified Bruhat--Tits tree by $T_{p^n}$.  It can be realized as a group quotient, $T_{p^n} = {\rm PGL}(2,\mathbb{Q}_{p^n}) / {\rm PGL}(2,\mathbb{Z}_{p^n})$, where $\mathbb{Z}_{p^n} = \mathbb{Z}_{\mathbb{Q}_{p^n}}$ \cite{Manin1976}.  We can use the same parametrization $(z_0,z)$ of points in $T_{p^n}$ as we did for $T_p$, and it is made precise by uniquely associating a point on the tree with the set $z + z_0 \mathbb{Z}_{p^n}$ of points on the boundary that can be reached by traveling upward from it.

Starting with the unramified field extension $\mathbb{Q}_{p^f}$, we may extend further by adding in $p^{1/e}$ for some $e>1$.  The residue field is still $\mathbb{F}_{p^f}$, and the natural uniformizer is $p^{1/e}$.  The Bruhat--Tits tree is unchanged from $T_{p^f}$, except in that the depth coordinate $z_0$ takes values $p^{m/e}$ for $m \in \mathbb{Z}$; thus it is a refinement of the Bruhat--Tits tree for $\mathbb{Q}_{p^f}$ in the sense of \cite{Ghoshal:2006zh}.  This ramified extension has two conflicting notions of dimensionality: The residue field $\mathbb{F}_{p^f}$ has dimension $f$ as a vector space over $\mathbb{F}_p$, but the full field extension has dimension $ef$ as a vector space over $\mathbb{Q}_p$.  Possibly the resulting holographic dynamics will show a mix of $f$-dimensional and $ef$-dimensional behaviors.

In the remainder of this paper we will stick with the simpler case of unramified extensions, where the conflicting notions of dimension described in the previous paragraph do not arise.  We will simplify notation by setting $q=p^n$ so that $\mathbb{Q}_q$ is the unramified extension of degree $n$.  We will abbreviate the norm $|\cdot|_{\mathbb{Q}_q}$ to $|\cdot|_q$ and the valuation $v_{\mathbb{Q}_q}$ to $v_q$.

\subsection{$p$-adic integration}

We will need to perform integrals over $\mathbb{Q}_p$ or over the unramified extension $\mathbb{Q}_q$.  Such integrals may be approximated as Riemann sums built by sampling a typical value of the integrand at each point in the Bruhat--Tits tree $T_q$, and then summing these typical values over all points at a fixed depth $z_0$.  In practice, we will often work with integrands which are piecewise constant over easily enumerated subsets of $\mathbb{Q}_q$, and then the integral can be replaced by a discrete sum.

In order to write down the Riemann sums of interest explicitly, it is helpful to define the sets
 \eqn{SmODef}{
  S^\omega_\mu \equiv \left\{ x \in \mathbb{Q}_q \colon 
    x = \sum_{m=\mu}^{\omega-1} a_m p^m \hbox{ where all $a_m \in \mathbb{F}_q$} 
    \right\}
 }
for $\omega > \mu$, and $S^\omega_\mu = \{0\}$ for $\omega \leq \mu$.  If $\omega>\mu$, then the elements of $S^\omega_\mu$ uniquely label the $q^{\omega-\mu}$ points in $T_q$ at depth $z_0 = p^\omega$ which can be accessed by going upward $\omega-\mu$ steps from the point $(p^\mu,0)$.  Given a function $f: \mathbb{Q}_q \to \mathbb{R}$ which is continuous and tends swiftly to $0$ when its argument is large in the $|\cdot|_q$ norm, we may approximate
 \eqn{fRiemannSum}{
  \int_{\mathbb{Q}_q} dx \, f(x) \approx \sum_{x \in S^\omega_\mu} q^{-\omega} f(x) \,,
 }
where we usually require $\omega > \mu$.  The approximate equality in \eno{fRiemannSum} becomes an exact equality in the limit where $\mu \to -\infty$ and $\omega \to \infty$.  We should think of $\mu$ as an infrared cutoff which essentially tells us to replace the integral over $\mathbb{Q}_q$ by an integral over $p^\mu \mathbb{Z}_q$.  The ultraviolet cutoff $\omega$ tells us to sample the integral at evenly spaced points, with a small volume $q^{-\omega}$ assigned to each point.  We observe that for fixed $\omega$ and fixed $\mu$ with $\omega \geq \mu$, we can split up $p^\mu \mathbb{Z}_q$ into many copies of $p^\omega \mathbb{Z}_q$, each shifted by an element of $S^\omega_\mu$.  Explicitly, we can write $p^\mu \mathbb{Z}_q$ as a disjoint union:
 \eqn{DisjointZ}{
  p^\mu \mathbb{Z}_q = \bigsqcup_{x \in S^\omega_\mu} \left( x + p^\omega \mathbb{Z}_q \right)
  \qquad\text{for fixed }\omega\geq\mu \,.
 }
The disjoint union \eno{DisjointZ} helps motivate the form of the right hand side of \eno{fRiemannSum}.  Using \eno{fRiemannSum}, one can show that
 \eqn{IntegerIntegral}{
  \int_{\mathbb{Z}_q} dx = 1 \,,
 }
and that
 \eqn{SimpleJacobian}{
  \int_{\xi S} dx = |\xi|_q^n \int_S dy
 }
for any measurable set $S \subset \mathbb{Q}_q$ and any fixed non-zero element $\xi \in \mathbb{Q}_q$.  Then $\xi S = \{ \xi s: s \in S\}$.  We think of the prefactor in \eno{SimpleJacobian} as coming from $|dx/dy|^n_q = |N(dx/dy)|_p$ where $x = \xi y$.  Since $\mathbb{U}_q = \mathbb{Z}_q \;\backslash\; p \mathbb{Z}_q$, we see by combining \eno{IntegerIntegral} and \eno{SimpleJacobian} that
 \eqn{UnitsIntegral}{
  \int_{\mathbb{U}_q} dx = 1 - {1 \over q} \,.
 }

In later sections we will need the Fourier transform over $\mathbb{Q}_q$.  The first ingredient is an additive character $\chi: \mathbb{Q}_q \to S^1$, where we think of $S^1$ as a complex phase.  The key properties of $\chi$ are $\chi(\xi+\eta) = \chi(\xi) \chi(\eta)$, $\chi(0) = 1$, and $\chi(\xi)^* = \chi(-\xi)$, where ${}^*$ means complex conjugation.  The standard additive character on $\mathbb{R}$ is $\chi(\xi) = {\rm e}^{2\pi i \xi}$.  On $\mathbb{Q}_p$, the standard choice is $\chi(\xi) = {\rm e}^{2\pi i [\xi]}$, where $[\xi]$ is the fractional part of $\xi$: That is, $[\xi] \in [0,1)$, and $\xi = [\xi] + m$ for some $m \in \mathbb{Z}_p$.  Note that $[\xi]$ is a rational number whose denominator is a power of $p$.  To handle an unramified extension $\mathbb{Q}_q$, we start with an arbitrary non-zero element
 \eqn{xiExpand}{
  \xi = \sum_{m=v_q(\xi)}^\infty b_m p^m \,,
 }
where each $b_m \in \mathbb{F}_q$ and $b_{v_q(\xi)} \neq 0$, and define the fractional part as $[\xi] = 0$ if $\xi \in \mathbb{Z}_q$, and
 \eqn{psiExtend}{
  [\xi] = \sum_{m=v_q(\xi)}^{-1} \Tr_{\mathbb{F}_q:\mathbb{F}_p}(b_m) p^m
 }
otherwise.  To make sense of \eno{psiExtend}, recall that $\Tr_{\mathbb{F}_q:\mathbb{F}_p}: \mathbb{F}_q \to \mathbb{F}_p$ is a homomorphism of the additive group structures on $\mathbb{F}_q$ and $\mathbb{F}_p$, and $\mathbb{F}_p$ can be identified with the set $\{0,1,2,\ldots,p-1\}$, so the right hand side of \eno{psiExtend} is a rational number in $[0,1)$ whose denominator is a power of $p$.  Now we can define $\chi(\xi) = {\rm e}^{2\pi i [\xi]}$ as in the case of $\mathbb{Q}_p$.

The Fourier and inverse Fourier transforms over $\mathbb{Q}_q$ can be defined as
 \eqn{FourierTransform}{
  f(x) = \int_{\mathbb{Q}_q} dk \, \chi(kx) \tilde{f}(k) \qquad\qquad
  \tilde{f}(k) = \int_{\mathbb{Q}_q} dx \, \chi(kx)^* f(x) \,.
 }
A key feature of the $p$-adic Fourier transform is that
 \eqn{GammaDef}{
  \gamma_q(x) \equiv 
   \left\{ \seqalign{\span\TR &\qquad\span\TT}{1 & for $x \in \mathbb{Z}_q$  \cr
    0 & otherwise} \right.
 }
is its own Fourier transform.  Indeed, if $k \in \mathbb{Z}_q$, then for all $x \in \mathbb{Z}_q$, we have $[kx] = 0$, so $\chi(kx) = 1$ and the second integral of \eno{FourierTransform} reduces to the integral of $1$ over $\mathbb{Z}_q$, which is $1$; whereas, if instead $k \notin \mathbb{Z}_q$, the character $\chi(kx)$ takes values symmetrically distributed around the unit circle in $\mathbb{C}$, so that the average is $0$.  Using \eno{SimpleJacobian} and \eno{GammaDef}, we can show that
 \eqn{FourierExamples}{
  \int_{\xi \mathbb{U}_q} dy \, \chi(y) = 
   |\xi|_q^n \left( \gamma_q(\xi) - {1 \over q} \gamma_q(p\xi) \right) = 
   \left\{ \seqalign{\span\TR &\qquad\span\TT}{
    |\xi|_q^n \left( 1 - {1 \over q} \right) & if $v_q(\xi) \geq 0$  \cr
    -1 & if $v_q(\xi) = -1$  \cr
    0 & if $v_q(\xi) < -1$\,.} \right.
 }
With the formula \eno{FourierExamples} in hand, we can find the Fourier transform of any function $f(x)$ which depends only on the norm $|x|_q$.

In the computation of two-point functions we will use
 \eqn{DeltaInt}{
  \int_{\mathbb{Q}_q} dx \, \chi(k_1 x) \chi(k_2 x) = \delta(k_1+k_2) \,.
 }
This formal relation is rendered meaningful by integrating both sides against a smooth function $\tilde{f}(k_1)$ which decreases rapidly for large argument:
 \eqn{fTest}{
  \int_{\mathbb{Q}_q} dk_1 \, \tilde{f}(k_1) 
   \int_{\mathbb{Q}_q} dx \, \chi(k_1 x) \chi(k_2 x) &= 
  \int_{\mathbb{Q}_q} dx \, \chi(k_2 x) 
   \int_{\mathbb{Q}_q} dk_1 \, \tilde{f}(k_1) \chi(k_1 x)  \cr &
   = \int_{\mathbb{Q}_q} dx \, \chi(k_2 x) f(x) = \tilde{f}(-k_2) \,,
 }
where in the first step we switched order of integrations (still as a formal manipulation), and the remaining steps are rigorously defined examples of Fourier transforms.

\section{Propagators}
\label{PROPAGATORS}

Following the philosophy of AdS/CFT, we would like to study a classical action on the modified Bruhat--Tits tree $T_q$, where $q=p^n$.  From the behavior of classical fields on this $p$-adic version of $\hbox{AdS}_{n+1}$, we expect to obtain correlators on the boundary, $\partial T_q = \mathbb{Q}_q$ (more precisely, $\partial T_q = \mathbb{P}^1(\mathbb{Q}_q)$).  Readers interested in the simplest examples may consistently set $n=1$ and $q=p$, so that the whole discussion reduces to the unmodified $(p+1)$-regular Bruhat--Tits tree and the boundary is just $\mathbb{Q}_p$ (more precisely, $\mathbb{P}^1(\mathbb{Q}_p)$).

\subsection{Action, bulk-to-bulk propagator, and mass formula}

Consider the following discrete bulk Euclidean action on the tree,
\eqn{BulkAction}{
S = \sum_{\langle ab \rangle} {1 \over 2} \left(\phi_a - \phi_b\right)^2 + \sum_a 
  \left( {1 \over 2} m_p^2 \phi_a^2 - J_a \phi_a \right) \,,
}
where $a$ and $b$ label vertices on the tree.  The notation $\langle ab \rangle$ indicates that the sum is over nearest neighboring lattice sites, or in other words over all edges of the tree.  The equation of motion derived from \eno{BulkAction} is
 \eqn{eom0}{
  (\square + m_p^2) \phi_a = J_a \,,
 }
where the laplacian on the tree is
 \eqn{TreeLap}{
  \square \phi_a = \sum_{\langle ab \rangle \atop \ a\ \rm fixed} (\phi_a - \phi_b) \,.
 }
The choice of sign in \eno{TreeLap} corresponds to the choice $\square = -{1 \over \sqrt{g}} \partial_\mu \sqrt{g} g^{\mu\nu} \partial_\nu$ on a real manifold, i.e.~$\square$ is positive definite as an operator acting on functions with swift fall-off at infinity.  A solution to \eno{eom0} is
 \eqn{PhiGreens}{
  \phi_a = \sum_b G(a,b) J_b \,,
 }
where the Green's function $G(a,b)$ satisfies
 \eqn{GreenDef}{
  (\square_a + m_p^2) G(a,b) = \delta(a,b) \,.
 }
In \eno{GreenDef} the laplacian $\square_a$ is understood to act on the first index of $G$, and
 \eqn{deltaDef}{
  \delta(a,b) = \left\{ \seqalign{\span\TR &\quad\span\TT}{1 & if $a = b$  \cr  0 & otherwise\,.}
    \right.
 }
A solution to \eno{GreenDef} depending only on the distance $d(a,b)$ between vertices $a$ and $b$ on the tree is
 \eqn{GabAnsatz}{
  G(a,b) = {\zetafct_p(2\Delta) \over p^\Delta} p^{-\Delta d(a,b)} \,,
 }
where $\Delta$ satisfies the relation
 \eqn{DeltaForm}{
  m_p^2 = -{1 \over \zetafct_p(\Delta-n) \zetafct_p(-\Delta)} \,.
 }
We have introduced the $p$-adic zeta functions
 \eqn{zetaForm}{
  \zetafct_p(s) \equiv {1 \over 1-p^{-s}} \,.
 }
If $m_p^2 = m_{BF,p}^2 \equiv -1/\zetafct_p(-n/2)^2$, then the unique real solution to \eno{DeltaForm} is $\Delta = n/2$.  If $m_p^2 > m_{BF,p}^2$, there are two real solutions $\Delta_\pm$ to \eno{DeltaForm} with $\Delta_+ + \Delta_- = n$ and $\Delta_+ > \Delta_-$.  If $m_{BF,p}^2 < m_p^2 < 0$, then $\Delta_\pm$ are both positive, whereas if $m_p^2 > 0$, $\Delta_+ > 0$ while $\Delta_- < 0$.  This is similar to the situation for real ${\rm AdS}_{n+1}$ of radius $L$, where $m_\infty^2 L^2 = \Delta (\Delta-n)$.  Following standard terminology, we will refer to constructions based on $\mathbb{R}$ as the ``Archimedean place.''\footnote{The Archimedean property of $\mathbb{R}$ is that if $a,b \in \mathbb{R}$ with $0 < |a| < |b|$, then for some $n \in \mathbb{Z}$ we have $|na| > |b|$.  By contrast, if $a,b \in \mathbb{Q}_p$ with $0 < |a|_p < |b|_p$, then since $|na|_p = |n|_p |a|_p \leq |a|_p$ for all $n \in \mathbb{Z}$, we have $|na|_p < |b|_p$ for all $n \in \mathbb{Z}$.  So the $p$-adic norm $|\cdot|_p$ is non-Archimedean.}  In either the Archimedean or $p$-adic places, we will assume $\Delta = \Delta_+ > n/2$ from here on and leave aside considerations of alternative quantization.

\subsection{Bulk-to-boundary propagator}

Next we want to formulate the bulk-to-boundary propagator $K(a,x)$, where $a \in T_q$ is a bulk point and $x \in \mathbb{Q}_q$ is a boundary point.  $K(a,x)$ should be a limit of $G(a,b)$ where $b$ is taken to the boundary point $x \in \mathbb{Q}_q$.  As this limit is taken with $a$ held fixed, we must multiply by some function of $b$ to keep the result finite while preserving the property
 \eqn{KCondition}{
  (\square_a + m_p^2) K(a,x) = 0
 }
throughout the bulk.  An obvious adaptation of the standard normalization convention is
 \eqn{KNorm}{
  \int_{\mathbb{Q}_q} dx \, K(z_0,z;x) = |z_0|_p^{n-\Delta} \,,
 }
where the power on the right hand side makes sense because then the right hand side itself is annihilated by $\square + m_p^2$, and we remember that the bulk point $a$ can be expressed as $(z_0,z)$.

We insist on translation invariance in the $\mathbb{Q}_q$ direction: that is, $K(z_0,z;x)$ depends on $z$ and $x$ only through the difference $z-x$.  We propose the form
 \eqn{Kexpress}{
  K(z_0,z;x) = {\zetafct_p(2\Delta) \over \zetafct_p(2\Delta-n)}
    {|z_0|_p^\Delta \over |(z_0,z-x)|_s^{2\Delta}} \,.
 }
By $|\cdot|_s$ we mean the supremum norm:
 \eqn{supNorm}{
  |(z_0,z-x)|_s = \sup\{|z_0|_p,|z-x|_q\} \,,
 }
where $|z_0|_p = p^{-\omega}$ when $z_0 = p^\omega$.

To check that \eno{Kexpress} is correct, it is enough to start by setting $z=0$ and holding $z_0$ fixed.  Then the bulk point is precisely the point along the trunk labeled by $z_0 = p^\omega$, which is to say the point at which the bush below $z_0 \mathbb{U}_q$ is rooted.  For points $x \in z_0 \mathbb{Z}_q = \bigsqcup_{m \geq \omega} p^m \mathbb{U}_q \sqcup \{0\}$, a path from $x$ to the bulk point $(z_0,0)$ goes straight down.  By contrast, to go from a boundary point $x \in p^{-1} z_0 \mathbb{U}_q$ to the bulk point $(z_0,0)$ we must go down to the root point $p^{-1} z_0$ and then back up one step to $z_0$.  This means there are two extra steps in the paths from points $x \in p^{-1} z_0 \mathbb{U}_q$ as compared to paths from points $x \in z_0 \mathbb{Z}_q$.  There are $2m$ extra steps if $x \in p^{-m} z_0 \mathbb{U}_q$, since we must go down to the root point $p^{-m} z_0$ and then back up $m$ steps to $z_0$.\footnote{The number of steps from a boundary point to a bulk point is always infinite, so the careful reader may prefer the more precise statement that if we start counting steps at some fixed depth $w_0$, with $|w_0|_p < |z_0|_p$, then the number of steps is the same for all $x \in z_0 \mathbb{Z}_q$, and increases by $2m$ for $x \in p^{-m} z_0 \mathbb{Z}_q$.}  Because $G(a,b) \propto p^{-\Delta d(a,b)}$, we are penalized by a factor of $p^{-2\Delta}$ for each extra step we take.  Hence $K(z_0,0;x)$ is constant for $x \in z_0 \mathbb{Z}_q$, whereas $K(z_0,0;x) = p^{-2\Delta m} K(z_0,0;0)$ when $|x|_q = p^m |z_0|_q$ for $m>0$.  Observing that
 \eqn{supNormExample}{
  |(z_0,x)|_s = \left\{ \seqalign{\span\TR & \qquad\span\TT}{
   |z_0|_p & for $|x|_q \leq |z_0|_p$  \cr
   p^m |z_0|_p & for $|x|_q = p^m |z_0|_p$ and $m>0$\,,} \right.
 }
we see that the factor $|(z_0,z-x)|_s^{2\Delta}$ in the denominator of the right hand side of \eno{Kexpress} is just what we need to account for the $x$-dependence of $K$.

To figure out the $z_0$ dependence of $K$, let's set $x=z=0$.  Then the number of steps from the boundary point to the bulk point (that is, from $x=0$ to the point marked $z_0 = p^\omega$ on the tree) increases by $1$ every time we decrease $\omega$ by $1$.  Each such step should decrease $K$ by a factor of $p^{-\Delta}$.  So we conclude that $K(z_0,0;0) \propto |z_0|_p^{-\Delta}$.  Together with the considerations of the previous paragraph, we see that $K(z_0,z;x) \propto |z_0|_p^\Delta/|(z_0,z-x)|_s^{2\Delta}$.  At this point one can explicitly verify the property \eno{KCondition}.

All that remains is to check that the overall normalization of $K$ in \eno{Kexpress} matches the condition \eno{KNorm}.  To this end we calculate the integral
 \eqn{IntegralDivided}{
  \int_{\mathbb{Q}_q} {dx \over |(z_0,x)|_s^{2\Delta}} &= 
   \int_{z_0 \mathbb{Z}_q} {dx \over |z_0|_p^{2\Delta}} + 
   \sum_{m=1}^\infty \int_{p^{-m} z_0 \mathbb{U}_q} {dx \over p^{2m\Delta} |z_0|_p^{2\Delta}}
     \cr 
   &= |z_0|_p^{n-2\Delta} \left[ 1 + \left( 1 - {1 \over q} \right) 
    \sum_{m=1}^\infty q^m p^{-2m\Delta} \right]  \cr
   &= |z_0|_p^{n-2\Delta} \left[ 1 + \left( 1 - {1 \over p^n} \right) \left(
     {1 \over 1 - p^{n-2\Delta}} - 1 \right) \right]  \cr
   &= |z_0|_p^{n-2\Delta} {\zetafct_p(2\Delta-n) \over \zetafct_p(2\Delta)} \,,
 }
where the key step was to split the integral over all of $\mathbb{Q}_q$ to integrals over disjoint domains across which the integrand is constant.  The final result in \eno{IntegralDivided} confirms the normalization in \eno{Kexpress}.

It is interesting to note that the normalized bulk-to-boundary propagator for a scalar field in Euclidean ${\rm AdS}_{n+1}$ is
 \eqn{Kreal}{
  K(z_0,\vec{z};\vec{x}) =
   {\zetafct_\infty(2\Delta) \over \zetafct_\infty(2\Delta-n)}
   {z_0^\Delta \over (z_0^2 + (\vec{z}-\vec{x})^2)^\Delta} \,,
 }
where we have used the normalization condition
 \eqn{RealKNorm}{
  \int_{\mathbb{R}^n} d^n x \, K(z_0,\vec{z};\vec{x}) = z_0^{n-\Delta}
 }
and introduced the local zeta function
 \eqn{zetaInfty}{
  \zetafct_\infty(s) = \pi^{-s/2} \Gammafct(s/2) \,.
 }

\subsection{Bulk-to-boundary propagator in Fourier space}

Recall in the real case that we may express
 \eqn{KfourierTransform}{
  K(z_0,\vec{z};\vec{x}) = \int_{\mathbb{R}^n} d^n k \, {\rm e}^{2\pi i \vec{k} \cdot (\vec{x}-\vec{z})}
    K(z_0,\vec{k}) \,,
 }
where
 \eqn{Kfourier}{
  K(z_0,\vec{k}) = {2 \over \zetafct_\infty(2\Delta-n)} k^{\Delta-{n \over 2}} z_0^{n \over 2}
    K_{\Delta-{n \over 2}}(2\pi kz_0) \,,
 }
where $K_\nu$ is a modified Bessel function of the second kind.  The positioning of the factors of $2\pi$ in \eno{KfourierTransform} is non-standard, but it is easily understood if $\vec{k}$ is thought of as momentum as the result of setting Planck's constant $h=1$ instead of the usual $\hbar=1$.  It is useful to note the asymptotics
 \eqn{Ksmall}{
  K(z_0,\vec{k}) = z_0^{n-\Delta} \left[ 1 + \ldots \right] + 
    {\zetafct_\infty(-2\Delta+n) \over \zetafct_\infty(2\Delta-n)} k^{2\Delta-n} z_0^\Delta
     \left[ 1 + \ldots \right]
 }
for small $kz_0$, while
 \eqn{Kbig}{
  K(z_0,\vec{k}) = {k^{\Delta-{n \over 2} - {1 \over 2}} z_0^{{n \over 2} - {1 \over 2}}
    \over \zetafct_\infty(2\Delta-n)} {\rm e}^{-2\pi kz_0} + \ldots
 }
for large $kz_0$.  In \eno{Ksmall}, $[1 + \ldots]$ denotes a Taylor series in $(kz_0)^2$.  Readers familiar with the standard formulas for bulk-to-boundary propagators in Fourier space may be amused to note that by using $\zetafct_\infty$ in favor of $\Gammafct$ and following the $h=1$ convention rather than $\hbar=1$, we obtain simpler expressions for the coefficients than the usual ones.

Now we would like to generalize \eno{Kfourier} to the $p$-adic case.  Explicitly,
 \eqn{PropFourier}{
  K(z_0,k) \equiv \int_{\mathbb{Q}_q} dx \, \chi(kx)^* K(z_0,0;x)
   = {\zetafct_p(2\Delta) \over \zetafct_p(2\Delta-n)} |z_0|_p^\Delta 
      \sum_{m \in \mathbb{Z}} \int_{p^m \mathbb{U}_q} dx 
        {\chi(kx)^* \over |(z_0,x)|_s^{2\Delta}} \,.
 }
In the second equality, we have split $\mathbb{Q}_q$ into nested spheres $p^m \mathbb{U}_q$.  We were able to drop integration over the point $0$ because this point has zero measure and the integrand is finite there.  $|(z_0,x)|_s^{2\Delta}$ depends on $x$ only through its norm $|x|_q$, so each integrand in the last expression in \eno{PropFourier} is constant over its domain $p^m \mathbb{U}_q$ of integration.  As a result we may use \eno{FourierExamples}.  After some work, we obtain the simple result
 \eqn{PFeval}{
  K(z_0,k) = \left( |z_0|_p^{n-\Delta} + |k|_q^{2\Delta-n} |z_0|_p^\Delta
     {\zetafct_p(-2\Delta+n) \over \zetafct_p(2\Delta-n)} \right) \gamma_q(kz_0) \,.
 }
(We could simplify \eno{PFeval} further using $\zetafct_p(-2\Delta+n) / \zetafct_p(2\Delta-n) = -p^{-2\Delta+n}$, but for comparison with \eno{Ksmall} it is best to leave \eno{PFeval} in terms of $p$-adic zeta functions.)  A remarkable point about \eno{PFeval} is that $K(z_0,k)$ is {\it exactly} a linear combination of the two power laws $|z_0|_p^{n-\Delta}$ and $|z_0|_p^\Delta$ down to the point where $|kz_0|_q = 1$, and then for larger $|kz_0|_q$ (meaning, further from the boundary of the Bruhat--Tits tree), $K(z_0,k)$ vanishes exactly, instead of the exponentially small behavior \eno{Kbig} observed in the real case.  The exact vanishing explains a point that may have been puzzling the reader: an on-shell scalar with momentum $k$ should take the form
 \eqn{PhiFourier}{
  \phi(z_0,z) = K(z_0,k) \chi(kz) \,,
 }
but how can this be a well-defined function on the Bruhat--Tits tree when $\chi(kz)$ itself is not?  The answer is that $\chi(kz)$ is well-defined at points on the tree $(z_0,z)$ precisely if $|kz_0|_q \leq 1$.  To see that this is true, remember that for a point $(z_0,z)$ on the tree, the $p$-adic coordinate $z$ can only be specified up to $O(z_0)$ corrections.  That means that $kz$ is determined up to $O(p^{v_q(kz_0)})$ corrections.  $\chi(kz)$ is well-defined if and only if the fractional part $[kz]$ is well-defined, which is the same as saying that $kz$ is determined up to $O(p^m)$ corrections for some $m \geq 0$.  We conclude that the condition for $\chi(kz)$ to be well-defined at $(z_0,z)$ is $v_q(kz_0) \geq 0$, and that is the same condition as $|kz_0|_q \leq 1$.  In other words, $\chi(kz)$ is not defined on the whole tree, but $\chi(kz) \gamma_q(kz_0)$ is, and that is enough for the expression \eno{PhiFourier} to be well-defined.

It is possible to go further and verify not only that \eno{PhiFourier} is well-defined, but also that it is a solution of $(\square+m_p^2) \phi = 0$.  There are four cases to consider:
 \begin{itemize}
  \item $|kz_0|_q > p$.  Trivial because $\phi(z_0,z) = \phi(pz_0,z) = \phi(z_0/p,z) = 0$ for all $z$ in this case.
  \item $|kz_0|_q < 1$.  In this case, $\phi$ is a sum of the two power laws permitted by the mass formula \eno{DeltaForm}, so the result is again trivial.
  \item $|kz_0|_q = p$.  This case is straightforward because $\phi(z_0,z) = \phi(z_0/p,z) = 0$ for all $z$, whereas for fixed $z$, $\phi(a)$ ranges over a non-trivial character of $\mathbb{F}_q$ as its argument $a$ ranges over the $q$ nearest neighbors of $(z_0,z)$ in the upward direction.
  \item $|kz_0|_q = 1$.  An explicit calculation starting from $(\square + m_p^2) \phi = 0$ leads to
 \eqn{LastCheck}{
  -q K(p|k|_q,k) + (q+1+m_p^2) K(|k|_q,k) = 0 \,,
 }
which is easily verified by direct substitution.
 \end{itemize}
In fact, using \eno{DeltaForm} one can show from \eno{LastCheck} that the relative coefficient between the two terms in parentheses in \eno{PFeval} must be as written there.  If we further require that $K(z_0,0) = |z_0|_p^{n-\Delta}$, which is the same as the normalization condition \eno{KNorm}, we can conclude that \eno{PFeval} is the only possible answer for $K(z_0,k)$, independently of the Fourier transform calculation given in \eno{PropFourier}.

\subsection{Cross-ratios and limiting procedures}
\label{CrossRatios}

\begin{figure}
  \centerline{\includegraphics[width=2.5in]{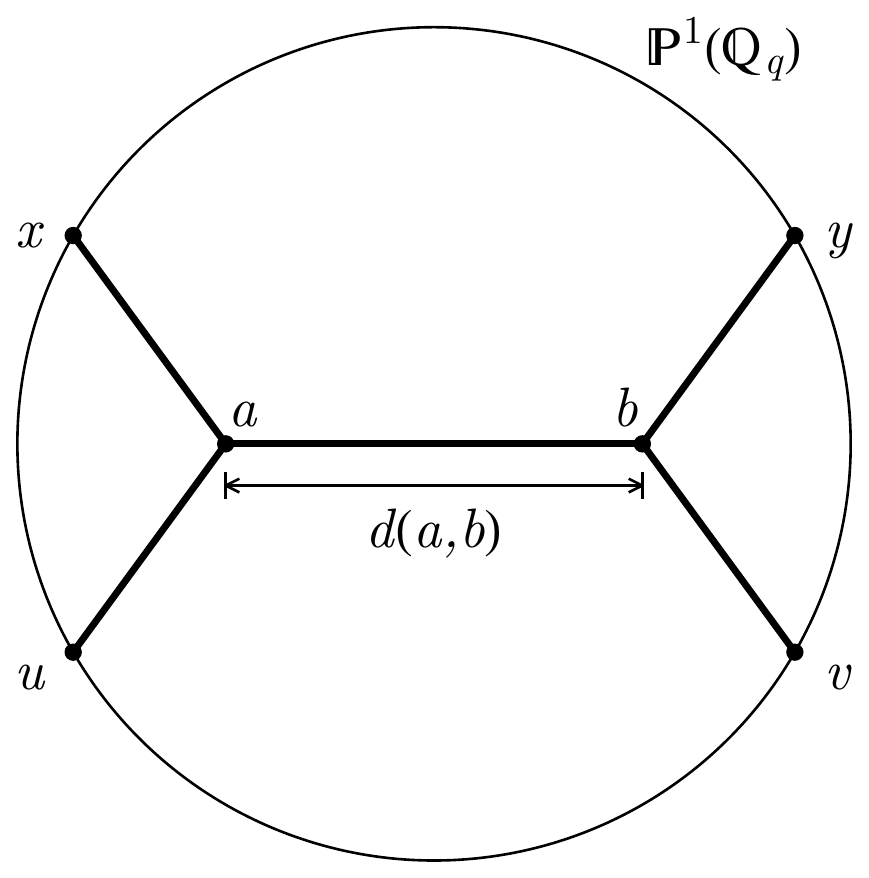}\qquad\qquad
   \includegraphics[width=2.6in]{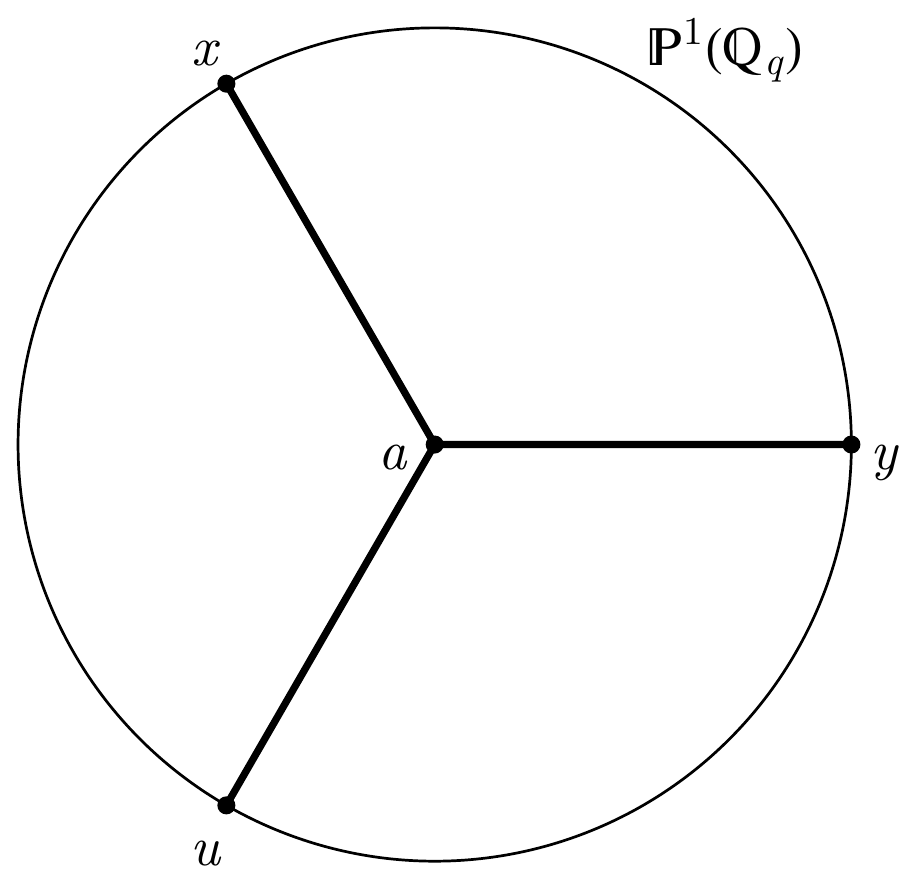}}
  \caption{Left: The distance $d(a,b)$ between $a$ and $b$ is the number of steps along $T_q$ in the path from $a$ to $b$.  The path from $x$ to $y$ in $T_q$ goes through $a$ and then $b$; likewise the path from $u$ to $v$.  The intersection of the paths from $x$ to $y$ and from $u$ to $v$ is precisely the path from $a$ to $b$.  Right: Paths on $T_q$ from three boundary points $x$, $y$, and $u$ meet at a unique bulk point $a$.}\label{dabCross}
 \end{figure}

It will sometimes be useful to have expressions for propagators which are less attached to the particular choice of depth coordinate, and correspondingly to a particular $\mathbb{Q}_q$ patch of the projective space $\mathbb{P}^1(\mathbb{Q}_q)$.  For this purpose, a key formula expresses the distance $d(a,b)$ between two points $a$ and $b$ on $T_q$ in terms of points $x$, $y$, $u$, and $v$ in $\mathbb{P}^1(\mathbb{Q}_q)$ such that the paths on $T_q$ from $x$ to $y$ and from $u$ to $v$ intersect precisely along the path from $a$ to $b$, as in figure~\ref{dabCross}:
 \eqn{CrossRatioDistance}{
  p^{-d(a,b)} = \left| {(x-u)(y-v) \over (x-y)(u-v)} \right|_q  = \left| {(x-u)(y-v) \over (x-v)(u-y)} \right|_q\,.
 }
In writing \eno{CrossRatioDistance}, we are assuming that none of $x$, $y$, $u$, and $v$ are at $\infty$.  The second and third expressions in \eno{CrossRatioDistance} are easily seen to be invariant under ${\rm PGL}(2,\mathbb{Q}_q)$.  If one of $x$, $y$, $u$, and $v$ is at $\infty$, then we must first apply a suitable ${\rm PGL}(2,\mathbb{Q}_q)$ transformation to all four points so that they all are mapped to $\mathbb{Q}_q$, and then use \eno{CrossRatioDistance}.

From \eno{GabAnsatz} and \eno{CrossRatioDistance} it is clear that we may express the bulk-to-bulk Green's function as
 \eqn{BulkBulkCross}{
  G(a,b) = {\zetafct_p(2\Delta) \over p^\Delta} 
   \left| {(x-u)(y-v) \over (x-y)(u-v)} \right|_q^\Delta \,.
 }
A little less obvious is the expression for the bulk-to-boundary propagator:
 \eqn{BulkBoundaryCross}{
  K(a,y) = {\zetafct_p(2\Delta) \over \zetafct_p(2\Delta-n)} 
    \left| {x-u \over (x-y)(u-y)} \right|_q^\Delta \,,
 }
where $a$ is the unique point where paths from $x$, $y$, and $u$ meet: see figure~\ref{dabCross}.  It is obvious that $K$ should be a limit of $G(a,b)$ as $b$ approaches the boundary: Explicitly,
 \eqn{KGlim}{
  K(a,y) = 2 \nu_p \lim_{v \to y} |y-v|_q^{-\Delta} G(a,b) \,,
 }
where $G(a,b)$ is written in the form \eno{BulkBulkCross} and $\nu_p$ is some constant.  The most straightforward way to obtain the prefactor written in \eno{BulkBoundaryCross} is to explicitly compare with \eno{Kexpress} using $y=0$, $x \in \mathbb{Q}_q^\times$, and $u \in \mathbb{Q}_q^\times$ with $|u|_q > |x|_q$.  Then, using the tall isosceles property of $|\cdot|_q$, we have from \eno{KGlim} the relation $K(a,y) = 2 \nu_p\frac{\zetafct_p(2\Delta)}{p^\Delta} |x|_q^{-\Delta}$, and from \eno{Kexpress} it is clear that $K(a,y) = {\zetafct_p(2\Delta) \over \zetafct_p(2\Delta-n)} |x|_q^{-\Delta}$.  The prefactor claimed in \eno{BulkBoundaryCross} follows immediately.

An alternative derivation of the prefactor in \eno{KGlim} proceeds by comparison with the Archimedean case, where in order to derive the relation analogous to \eno{KGlim} together with the correct prefactor, the natural starting point is the relation
 \eqn{KGderivativeReals}{
 K(z_0,\vec{z};\vec{x}) = \lim_{x_0 \rightarrow 0} \sqrt{h}\, x_0^{\Delta_-} n\cdot \partial G(z_0, \vec{z}; x_0, \vec{x})\,,
 }  
where $h$ is the determinant of the induced metric and $n$ is the outward facing normal vector at the boundary $x_0=0$.  (The relation \eno{KGderivativeReals} is essentially in the spirit of the treatment of \cite{Muck:1998rr}.)  Here we are using the short-hand $\Delta_- = n- \Delta_+ = n-\Delta$. Then using Green's second identity
 \eqn{GreenIdentity}{
 \int_\mathcal{M} d^{n+1}x \sqrt{g} \left(\phi(\square + m_{\infty}^2)\psi - \psi (\square + m_{\infty}^2) \phi \right) = -\int_{\partial \mathcal{M}} d^n y \sqrt{h} \left(\phi n \cdot \partial \psi - \psi n \cdot \partial \phi \right)
 }
with the substitutions $\phi(z_0,\vec{z}) = G(z_0,\vec{z};x_0,\vec{x})$ and $\psi(z_0,\vec{z}) = K(z_0,\vec{z};\vec{x})$, together with the scalings near the boundary
 \eqn{ScalingsReals}{
 z_0 \partial_{z_0} G(z_0,\vec{z};x_0,\vec{x}) = \Delta_+ G(z_0,\vec{z};x_0,\vec{x}) \qquad \qquad \lim_{z_0 \rightarrow 0} K(z_0,\vec{z};\vec{x}) = z_0^{\Delta_-} \delta(\vec{z}-\vec{x})
  \,,
 }
it follows that 
 \eqn{KGlimReals}{ 
 K(z_0,\vec{z};\vec{x}) = 2 \nu_\infty \lim_{x_0 \rightarrow 0}   x_0^{-\Delta_+} G(z_0,\vec{z}; x_0, \vec{x}) \qquad\hbox{where}\qquad 2\nu_\infty \equiv \Delta_+ - \Delta_- = 2\Delta -n\,.
 }
In the $p$-adics, to derive \eno{KGlim} together with the coefficient $\nu_p$, it is more convenient to start instead with the identity
 \eqn{PartialIntegration}{
 \phi_a = \sum_{b \in T_q} \left(\phi_b \left(\square_b + m_p^2\right) G(a,b) - G(a,b) \left(\square_b + m_p^2\right) \phi_b \right)
 }
where $a=(w_0,w)$ and $b=(z_0,z)$ are vertices on $T_q$, and rewrite $\phi_a$ on the l.h.s.~using
 \eqn{phiRiemannSum}{
 \phi_a = \int_{\mathbb{Q}_q} dz\, K(a,z) \phi_0(z) \approx \sum_{z\in S_\mu^\Omega} q^{-\Omega} K(a,z) \phi_0(z)\,,
 }
where $\mu$ and $\Omega$ are the infrared and ultraviolet cutoffs, respectively (see the discussion following \eno{fRiemannSum}), chosen such that $p^\mu < |a|_q < p^\Omega$. With some work, the r.h.s.~of \eno{PartialIntegration} can be partially integrated, and using the scalings
 \eqn{ScalingsPadics}{
 G(a;z_0/p,z)\Big|_{z_0=p^\Omega} = p^{\Delta_+} G(a;z_0,z)\Big|_{z_0=p^\Omega} \qquad \qquad \lim_{|z_0|_p \rightarrow 0} \phi(z_0,z) = |z_0|_p^{\Delta_-} \phi_0(z)\,,
 }
we arrive at
 \eqn{KGlimPadics}{
 K(z_0,z;x) = 2\nu_p \lim_{|x_0|_p \rightarrow 0} |x_0|_p^{-\Delta_+} G(z_0,z; x_0,x)
 }
where 
 \eqn{nupForm}{
  2\nu_p \equiv p^{\Delta_+} - p^{\Delta_-} = {p^\Delta \over \zetafct_p(2\Delta -n)}\,.
 } 
Using \eno{nupForm}, the final result of taking the limit \eno{KGlimPadics} agrees with \eno{Kexpress} and \eno{BulkBoundaryCross}.

When computing correlators, it will sometimes be convenient to refer to the unnormalized propagators
 \eqn{KGunnormalized}{
  \hat{G}(a,b) &\equiv \left| {(x-u)(y-v) \over (x-y)(u-v)} \right|_q^\Delta 
    = p^{-\Delta d(a,b)}  \cr
  \hat{K}(a,y) &\equiv \left| {x-u \over (x-y)(u-y)} \right|_q^\Delta 
    = {|z_0|_p^\Delta \over |(z_0,z-y)|_s^{2\Delta}} \,,
 }
where in the first line the arrangement of $a$, $b$, $x$, $y$, $u$, and $v$ are as described around \eno{CrossRatioDistance}, and in the second line $a = (z_0,z)$ is the unique point where paths from $x$, $y$, and $u$ meet.

\section{Correlators}
\label{CORRELATORS}

Let's start with a naive approach to $p$-adic AdS/CFT two-point correlator, which misses some overall factors but nevertheless give us some interesting partial guidance on what to expect.  In this naive approach, the two-point function is extracted as the limit of the bulk-to-boundary propagator, where the bulk point is taken to the boundary.  Explicitly, starting from \eno{Kexpress},
 \eqn{KeXlim}{
  \langle {\cal O}(x) {\cal O}(y) \rangle_{p,\rm naive} &= 
    \lim_{x_0 \to 0} |x_0|_p^{-\Delta} K(x_0,x;z) = 
    {\zetafct_p(2\Delta) \over \zetafct_p(2\Delta-n)} \lim_{x_0 \to 0}
      {1 \over |(x_0,x-y)|_s^{2\Delta}}  \cr
    &= {\zetafct_p(2\Delta) \over \zetafct_p(2\Delta-n)} {1 \over |x-y|_q^{2\Delta}} \,,
 }
where $\lim_{x_0 \to 0}$ refers to setting $x_0 = p^\omega$ and sending $\omega \to +\infty$ so that $x_0$ becomes small in the $p$-adic norm $|\cdot|_p$.  (The same answer could be obtained by starting from \eno{BulkBoundaryCross}, multiplying by $|x-u|_q^{-\Delta}$, and taking the limit $u \to x$ in the topology of $\mathbb{Q}_q$.)  An equally naive calculation in the Archimedean case starts with \eno{Kreal} and works the same way:
 \eqn{KrXlim}{
  \langle {\cal O}(x) {\cal O}(y) \rangle_{\infty,\rm naive} = 
    \lim_{x_0 \to 0} x_0^{-\Delta} K(x_0,\vec{x};\vec{y}) = 
    {\zetafct_\infty(2\Delta) \over \zetafct_\infty(2\Delta-n)} 
     {1 \over |\vec{x}-\vec{y}|^{2\Delta}} \,.
 }
We can add a bit of formal polish in the unextended case $n=1$ by noting that if $x$ and $y$ are rational, then we can define
 \eqn{OOv}{
  \langle {\cal O}(x) {\cal O}(y) \rangle_{v,\rm naive} = 
    {\zetafct_v(2\Delta) \over \zetafct_v(2\Delta-1)} {1 \over |x-y|_v^{2\Delta}}
 }
equally for $v=p$ and $v=\infty$, where $|\cdot|_\infty$ is the ordinary absolute value.  We are led to the adelic relation
 \eqn{OOadelic}{
  \langle {\cal O}(x) {\cal O}(y) \rangle_{\mathbb{A},\rm naive} \equiv
   \prod_v \langle {\cal O}(x) {\cal O}(y) \rangle_{v,\rm naive} = 
    {\zetafct_{\mathbb{A}}(2\Delta) \over \zetafct_{\mathbb{A}}(2\Delta-1)} \qquad
    \hbox{for unequal $x,y \in \mathbb{Q}$\,,}
 }
where the product is over all primes as well as $\infty$, and we used the key relation
 \eqn{AdelicProduct}{
  \prod_v |\xi|_v = 1 \qquad\hbox{for $\xi \in \mathbb{Q}$\,.}
 }
We have also introduced the adelic zeta function,
 \eqn{zetaComplete}{
  \zetafct_{\mathbb{A}}(s) \equiv \prod_v \zetafct_v(s) = \pi^{-s/2} \Gammafct(s/2) \zetafct(s) \,,
}
where $\zetafct(s) = \sum_{n=1}^\infty {1 \over n^s}$ is the ordinary Riemann zeta function.  The adelic zeta function obeys the simple functional relation $\zetafct_{\mathbb{A}}(s) = \zetafct_{\mathbb{A}}(1-s)$, and its non-trivial zeros are the same as the non-trivial zeros of $\zetafct(s)$, i.e.~a discrete sequence at $\Re s = 1/2$ according to the Riemann Hypothesis.

The above treatment of the two-point correlator is wrong (or, at least, against the usual spirit of AdS/CFT) because it leaves the on-shell action entirely out of the story.  In general---up to subtleties with regularization as discussed in the next section---the correct expression for $p$-adic correlators is
 \eqn{pGreenAgain}{
  -\log \left\langle \exp\left\{ \int_{\mathbb{Q}_q} dz \, \phi_0(z) {\cal O}(z)
      \right\} \right\rangle_p
    = \extremum_{\phi \to \phi_0} S[\phi]
 }
where by $\phi \to \phi_0$ we mean
 \eqn{pPhiApproach}{
  \lim_{z_0 \to 0} |z_0|_p^{\Delta-n} \phi(z_0,z) = \phi_0(z) \,,
 }
and $S[\phi]$ is the bulk action~\eno{BulkAction} or some generalization thereof, for example
 \eqn{Snonlinear}{
  S[\phi] = \eta_p \sum_{\langle ab \rangle} {1 \over 2} (\phi_a-\phi_b)^2 + 
   \eta_p \sum_a \left( {1 \over 2} m_p^2 \phi_a^2 + 
    {g_3 \over 3!} \phi_a^3 + {g_4 \over 4!} \phi_a^4
     \right) \,,
 }
where $\eta_p$, $g_3$, and $g_4$ are constants.  The formula \eno{pGreenAgain} is closely analogous to the standard AdS/CFT prescription of \cite{Gubser:1998bc,Witten:1998qj}, and it should be understood as receiving corrections from loops in the bulk, so that the full story is that the partition functions of the bulk and boundary coincide when appropriately sourced.

\subsection{Two-point function}
\label{TwoPoint}

As a warmup to $p$-adic calculations, let's review the standard account for Archimedean ${\rm AdS}_{n+1}$, using Fourier space since it's easier to sort out prefactors reliably in Fourier space than in position space.  The on-shell scalar configuration we are interested in is
 \eqn{phiSoln}{
  \phi(z_0,\vec{z}) = \lambda_1 {\rm e}^{2\pi i \vec{k}_1 \cdot \vec{z}} K_\epsilon(z_0,\vec{k}_1) + 
    \lambda_2 {\rm e}^{2\pi i \vec{k}_2 \cdot \vec{z}} K_\epsilon(z_0,\vec{k}_2) \,.
 }
We have defined
 \eqn{KepsDef}{
  K_\epsilon(z_0,\vec{k}) \equiv {K(z_0,\vec{k}) \over K(\epsilon,\vec{k})} = 
    {z_0^{n-\Delta} + \zeta_R k^{2\Delta-n} z_0^\Delta \over
     \epsilon^{n-\Delta} + \zeta_R k^{2\Delta-n} \epsilon^\Delta} + \ldots 
     \qquad\hbox{where}\qquad
     \zeta_R \equiv {\zetafct_\infty(-2\Delta+n) \over \zetafct_\infty(2\Delta-n)} \,.
 }
The notation ${} + \ldots$ in the third expression of \eno{KepsDef} reminds us that we have dropped terms which are subleading to the ones shown in both the numerator and the denominator by positive even powers of $kz_0$ or $k\epsilon$.  The point of \eno{KepsDef} is that we have arranged to have $\phi(\epsilon,\vec{z}) = \lambda_1 {\rm e}^{2\pi i \vec{k}_1 \cdot \vec{z}} + \lambda_2 {\rm e}^{2\pi i \vec{k}_2 \cdot \vec{z}}$, which we use as source for a regulated version of the operator of interest, call it ${\cal O}_\epsilon$.  Then the prescription we will use for Green's functions is
 \eqn{GreenPrescription}{
  -\log \left\langle \exp\left\{
    \int_{\mathbb{R}^n} d^n z \, \phi_\epsilon(\vec{z}) {\cal O}_\epsilon(\vec{z}) 
    \right\} \right\rangle_\infty
   = \extremum_{\phi(\epsilon,\vec{z}) = \phi_\epsilon(\vec{z})} S_\epsilon[\phi] \,,
 }
where
 \eqn{Sreal}{
  S_\epsilon[\phi] = \eta_\infty \int_{z_0 > \epsilon} d^{n+1} z \, \sqrt{\det g_{\mu\nu}} 
   \left[ {1 \over 2} g^{\mu\nu} \partial_\mu \phi \partial_\nu \phi + 
    {1 \over 2} m_\infty^2 \phi^2 \right] \,,
 }
and the ${\rm AdS}_{n+1}$ metric is $ds^2 = g_{\mu\nu} dz^\mu dz^\nu = {L^2 \over z_0^2} \left( dz_0^2 + d\vec{z}^2 \right)$.  The prefactor $\eta_\infty$ is related to the gravitational coupling in string theory realizations of AdS/CFT.  In order to make it easy to compute the extremum, we add to the action a multiple of the equation of motion:
 \eqn{Sonshell}{
  S_\text{on-shell} &= \eta_\infty \int_{z_0 > \epsilon} d^{n+1} z \, \sqrt{\det g_{\mu\nu}} \left[
   {1 \over 2} g^{\mu\nu} \partial_\mu \phi \partial_\nu \phi + 
    {1 \over 2} m_\infty^2 \phi^2 - {1 \over 2} \phi (\square + m_\infty^2) \phi \right]  \cr
   &= -{\eta_\infty \over 4} \int_{z_0 > \epsilon} d^{n+1} z \, \sqrt{\det g_{\mu\nu}}
    \square \phi^2  \cr
   &= -{\eta_\infty \over 4} \left( {L \over \epsilon} \right)^{n-1} 
     \int_{z_0 = \epsilon} d^n z \, \partial_{z^0} \phi^2 \,,
 }
where $\square = -{1 \over \sqrt{g}} \partial_\mu \sqrt{g} g^{\mu\nu} \partial_\nu$.  In order to extract the desired two-point function, we compute
 \eqn{TwoPointFunction}{
  \langle {\cal O}_\epsilon(\vec{k}_1) {\cal O}_\epsilon(\vec{k}_2) \rangle_\infty &= 
   \left. -{\partial^2 S_\text{on-shell} \over \partial\lambda_1 \partial\lambda_2} 
    \right|_{\lambda_1=\lambda_2 = 0}  \cr
   &= {\eta_\infty \over 2} \left( {L \over \epsilon} \right)^{n-1} \left(
     \int_{\mathbb{R}^n} d^n z \, {\rm e}^{2\pi i (\vec{k}_1+\vec{k}_2) \cdot \vec{z}} \right)
     \left[ \partial_{z_0} \left( K_\epsilon(z_0,\vec{k}_1) K_\epsilon(z_0,\vec{k}_2) \right)
      \right]_{z_0 = \epsilon}  \cr
   &= \eta_\infty {L^{n-1} \over \epsilon^n} 
    \delta(\vec{k}_1 + \vec{k}_2) \left[ -\Delta+n + 
      (2\Delta-n) (k_1 \epsilon)^{2\Delta-n} \zeta_R + \ldots \right] \,.
 }
We discard the $k_1$-independent term from inside square brackets in the last expression of \eno{TwoPointFunction} on grounds that its Fourier transform is a pure contact term in position space.  The terms we have omitted by writing $\ldots$ inside square brackets are suppressed by positive integer powers of $(k_1\epsilon)^{2\Delta-n}$ relative to the last term shown, so for small $\epsilon$ and fixed $k_1$ we may discard them too.  (We are ignoring the possibility of alternative quantization.)  Thus, if we ignore contact terms and also drop terms subleading in $\epsilon$, we wind up with
 \eqn{TPFposition}{
  \langle {\cal O}_\epsilon(\vec{x}_1) {\cal O}_\epsilon(\vec{x}_2) \rangle_\infty &= 
    \int_{\mathbb{R}^n} d^n k_1 d^n k_2 \, 
      {\rm e}^{2\pi i (\vec{k}_1 \cdot \vec{x}_1 + \vec{k}_2 \cdot \vec{x}_2)}
      \langle {\cal O}_\epsilon(\vec{k}_1) {\cal O}_\epsilon(\vec{k}_2) \rangle_\infty  \cr
     &= \eta_\infty L^{n-1} \epsilon^{2(\Delta-n)} (2\Delta-n) \zeta_R 
       \int_{\mathbb{R}^n} d^n k_1 \, {\rm e}^{2\pi i \vec{k}_1 \cdot \vec{x}_{12}}
        k_1^{2\Delta-n}  \cr
     &= \eta_\infty L^{n-1} \epsilon^{2(\Delta-n)} 
         (2\Delta-n) {\zetafct_\infty(2\Delta) \over \zetafct_\infty(2\Delta-n)} 
         {1 \over |\vec{x}_{12}|^{2\Delta}} \,,
 }
where we have set $\vec{x}_{12} = \vec{x}_1 - \vec{x}_2$.  In the last equality of \eno{TPFposition} we started with \eno{Kreal} and \eno{Ksmall} and expanded at small $z_0$ to obtain
 \eqn{FourierInt}{
  \zeta_R \int_{\mathbb{R}^n} d^n k \, {\rm e}^{2\pi i \vec{k} \cdot \vec{x}} k^{2\Delta-n}
    = {\zetafct_\infty(2\Delta) \over \zetafct_\infty(2\Delta-n)} {1 \over |\vec{x}|^{2\Delta}} \,.
 }
Note that the precise value of $\zeta_R$ doesn't matter, since it enters the holographic calculation \eno{TwoPointFunction} from the Fourier-space bulk-to-boundary propagator, and from precisely the same propagator we can extract \eno{FourierInt}.  The overall normalization of this propagator {\it does} enter, and it leads to the $\zetafct_\infty(2\Delta)/\zetafct_\infty(2\Delta-n)$ factor in the last expression of \eno{TPFposition}.  The factor $2\nu_\infty = 2\Delta-n$ arises when passing from the second line of \eno{TwoPointFunction} to the third line: That is, it is related to evaluating the $z_0$ derivative of $\phi^2$ at $z_0 = \epsilon$.  This is similar to the way the normal derivative in \eno{KGderivativeReals} leads to a factor of $2\nu_\infty$ in \eno{KGlimReals}.  We will therefore refer to $2\Delta-n$ as a boundary factor.  To obtain the final form of the two-point function, we note that $K(\epsilon,\vec{k}) \approx \epsilon^{n-\Delta}$ for small $\epsilon$.  Thus the solution $\phi(z_0,\vec{z})$ in \eno{phiSoln} contains an extra factor of $\epsilon^{\Delta-n}$, which can be regarded as a leg factor for defining a truly local operator:
 \eqn{OlocDef}{
  {\cal O}(\vec{x}) = \lim_{\epsilon \to 0} \epsilon^{n-\Delta} {\cal O}_\epsilon(\vec{x}) \,.
 }
Using \eno{OlocDef}, we obtain the final answer
 \eqn{OOfinal}{
  \langle {\cal O}(\vec{x}_1) {\cal O}(\vec{x}_2) \rangle_\infty = 
    \eta_\infty L^{n-1} (2\Delta-n) {\zetafct_\infty(2\Delta) \over \zetafct_\infty(2\Delta-n)}
      {1 \over |\vec{x}_{12}|^{2\Delta}} \,.
 }
This expression is valid only up to contact terms.  The more general expression for fully local correlators is
 \eqn{GreenAgain}{
  -\log \left\langle \exp\left\{ \int_{\mathbb{R}^n} d^n z \, \phi_0(\vec{z}) {\cal O}(\vec{z})
      \right\} \right\rangle_\infty
    = \extremum_{\phi \to \phi_0} S[\phi]
 }
where by $\phi \to \phi_0$ we mean
 \eqn{PhiApproach}{
  \lim_{z_0 \to 0} z_0^{\Delta-n} \phi(z_0,\vec{z}) = \phi_0(\vec{z}) \,,
 }
and $S[\phi]$ is the same as $S_\epsilon[\phi]$ in \eno{Sreal} but integrated over all of ${\rm AdS}_{n+1}$.

Now let's consider the analogous computation on the Bruhat--Tits tree, using the action \eno{Snonlinear}.  Only the quadratic terms are of interest to us since, for now, we only want the two-point function and are not concerned with loop corrections.  The on-shell scalar configuration of interest is
 \eqn{pphiSoln}{
  \phi(z_0,z) = \lambda_1 \chi(k_1 z) K_\epsilon(z_0,k_1) + 
    \lambda_2 \chi(k_2 z) K_\epsilon(z_0,k_2) \,.
 }
We have defined
 \eqn{pKepsDef}{
  K_\epsilon(z_0,k) \equiv {|z_0|_p^{n-\Delta} + \zeta_R |k|_q^{2\Delta-n} |z_0|_p^\Delta \over
   |\epsilon|_p^{n-\Delta} + \zeta_R |k|_q^{2\Delta-n} |\epsilon|_p^\Delta} \gamma_q(kz_0)
 }
where now
 \eqn{zetaRpadic}{
  \zeta_R \equiv {\zetafct_p(-2\Delta+n) \over \zetafct_p(2\Delta-n)} = -p^{-2\Delta+n} \,.
 }
Note that we cannot quite follow \eno{KepsDef} because $K(z_0,k) / K(\epsilon,k)$ is ill-defined for $|\epsilon k|_q > 1$.  In practice, we aim to keep momenta fixed while we take a limit $\epsilon \to 0$ (in the $p$-adic sense), so we will never encounter a situation where $|\epsilon k|_q > 1$.  As compared to \eno{KepsDef}, it is notable that in \eno{pKepsDef} we are not discarding any subleading terms at all.  This makes the structure of contact terms simpler.  Note that $\phi(\epsilon,z) = \lambda_1 \chi(k_1 z) + \lambda_2 \chi(k_2 z)$ (on the assumption $|\epsilon k|_q \leq 1$), so the obvious adaptation of \eno{GreenPrescription} is
 \eqn{pGreenPrescription}{
  -\log \left\langle \exp\left\{ \int_{\mathbb{Q}_q} dz \, \phi_\epsilon(z)
    {\cal O}_\epsilon(z) \right\} \right\rangle_p = 
   \extremum_{\phi(\epsilon,z) = \phi_\epsilon(z)} S_\epsilon[\phi]
 }
where
 \eqn{Spadic}{
  S_\epsilon[\phi] = \eta_p \sum_{|\epsilon|_p < |a_0|_p} \left[ 
    {1 \over 4} \sum_{\langle ab \rangle \atop a\ \rm fixed} (\phi_a-\phi_b)^2 + 
    {1 \over 2} m_p^2 \phi_a^2 \right] \,,
 }
where in a slight abuse of notation we use $a_0$ to mean the value of the depth coordinate $z_0$ at the point $a \in T_q$.  The reader should be forewarned that the cutoff procedure has an $O(1)$ impact on the normalization of the two-point function.  We are following what seems like the most sensible approach in \eno{Spadic} of first writing the sum $\sum_{\langle ab \rangle}$ over edges as a sum over vertices with an inner sum over the edges coming off of each vertex, and then restricting only the outer sum over vertices.  We have chosen to restrict the sum to points $a$ with $|\epsilon|_p < |a_0|_p$; later we will consider what happens if we say instead $|\epsilon|_p \leq |a_0|_p$.  It would be interesting to explore more systematically the full range of possible cutoffs for the sum.

The next step is to reduce the sum \eno{Spadic} to a boundary term.  Toward this end, we note
 \eqn{AddEom}{
  {1 \over 4} \sum_{\langle ab \rangle \atop a\ \rm fixed} (\phi_a-\phi_b)^2 + 
    {1 \over 2} m_p^2 \phi_a^2 - {1 \over 2} \phi_a (\square + m_p^2) \phi_a = 
    -{1 \over 4} \square \phi_a^2
 }
Thus, by adding a multiple of the equation of motion to the action \eno{Spadic}, we obtain
 \eqn{Stelescoping}{
  S_\text{on-shell} = -{\eta_p \over 4} 
   \sum_{|\epsilon|_p < |a_0|_p \leq |M|_p \atop \rm restricted} \square \phi_a^2 \,,
 }
where we have imposed an infrared cutoff by restricting the sum to run over only those points in the subtree rooted at $p^{v_q(M)}$.  To simplify notation, it helps to consistently set
 \eqn{zMeps}{
  z_0 = p^\omega \qquad M = p^\mu \qquad \epsilon = p^\Omega \,,
 }
with $\mu$ large and negative while $\Omega$ is large and positive.  Denoting a point $a \in T_q$ by $a = (z_0,z)$, we can enumerate the points in the sum \eno{Stelescoping} first by letting $\omega$ run over the integers in $[\mu,\Omega)$ and then, for each fixed $\omega$, letting $z$ run over $S^\omega_\mu$.  Next we need to have an explicit way of labeling the points $b$ which are the nearest neighbors of $a$.  Writing
 \eqn{zExplicit}{
  z = \sum_{m=\mu}^{\omega-1} \kappa_m p^m \in S^\omega_\mu \qquad\hbox{where each
    $\kappa_m \in \mathbb{F}_q$\,,}
 }
we see that
 \eqn{zDown}{
  z \to [z]_{\omega-1} \equiv \sum_{m=\mu}^{\omega-2} \kappa_m p^m
 }
is a $q$-to-$1$ map from $S^\omega_\mu$ to $S^{\omega-1}_\mu$ provided $\omega > \mu$, and if $\omega \leq \mu$ it is the trivial $1$-to-$1$ map---since in this latter case $S^\omega_\mu = S^{\omega-1}_\mu = \{0\}$.  Note that for $\mathbb{Q}_p$, $[z]_0$ is just the fractional part of $z$.  The map $(z_0,z) \to (z_0/p,[z]_{\omega-1})$ takes a point $(z_0,z)$ to its nearest neighbor in the downward direction (i.e.~the nearest neighbor one step closer to $\infty$).  We can also define $q$ $1$-to-$1$ maps from $S^\omega_\mu$ to $S^{\omega+1}_\mu$ as follows:
 \eqn{zUp}{
  z \to z + p^\omega \kappa
 }
where $\kappa \in \mathbb{F}_q$.  Then the maps $(z_0,z) \to (pz_0,z + p^\omega \kappa)$ take a point $(z_0,z)$ to its nearest neighbors in the upward direction.  Now we can rewrite \eno{Stelescoping} as
 \eqn{StExplicit}{
  S_\text{on-shell} &= -{\eta_p \over 4} \sum_{\omega=\mu}^{\Omega-1}
   \sum_{z \in S^\omega_\mu} \left[ (q+1) \phi(p^\omega,z)^2 - 
    \phi(p^{\omega-1},[z]_{\omega-1})^2 - 
    \sum_{\kappa \in \mathbb{F}_q} \phi(p^{\omega+1},z+p^\omega \kappa)^2 \right]  \cr
   &= -{\eta_p \over 4} \left[ \sum_{\omega=\mu}^{\Omega-1} 
     \sum_{z \in S^\omega_\mu} (q+1) \phi(p^\omega,z)^2 - 
     \sum_{\omega=\mu}^{\Omega-2} \sum_{z \in S^\omega_\mu} q \phi(p^\omega,z)^2 -
       \phi(p^{\mu-1},0)^2 \right.  \cr &\qquad\qquad\qquad\left. {} - 
     \sum_{\omega=\mu+1}^\Omega \sum_{z \in S^\omega_\mu} \phi(p^\omega,z)^2 \right]  \cr
   &= -{\eta_p \over 4} \left[ \sum_{z \in S^{\Omega-1}_\mu} q \phi(p^{\Omega-1},z)^2 - 
     \sum_{z \in S^\Omega_\mu} \phi(p^\Omega,z)^2 + \phi(p^\mu,0)^2 - \phi(p^{\mu-1},0)^2
     \right] \,.
 }
Because the Fourier space propagator vanishes identically for sufficiently large $|kz_0|_q$, we can drop the last two terms in square brackets in the last line of \eno{StExplicit}.  The resulting expression is the discrete version of the last line of \eno{Sonshell}.

We now compute the two-point function as
 \eqn{pTwoPointFunction}{
  \langle {\cal O}_\epsilon(k_1) {\cal O}_\epsilon(k_2) \rangle_p &= 
     -{\partial^2 S_\text{on-shell} \over \partial\lambda_1 \partial\lambda_2}  \cr
   &= {\eta_p \over 2} \left( q \sum_{z \in S_{\mu}^{\Omega-1}} \chi((k_1+k_2)z) \right)  \cr
   &\qquad\qquad{} 
     \times \left[ p^{2n-2\Delta} {1 + \zeta_R p^{2\Delta-n} |k_1 \epsilon|_q^{2\Delta-n} \over
       1 + \zeta_R |k_1 \epsilon|_q^{2\Delta-n}}
       {1 + \zeta_R p^{2\Delta-n} |k_2 \epsilon|_q^{2\Delta-n} \over
       1 + \zeta_R |k_2 \epsilon|_q^{2\Delta-n}} \right]  \cr
   &\qquad{} - {\eta_p \over 2} \left( \sum_{z \in S_{\mu}^\Omega} 
     \chi((k_1+k_2)z) \right) \,.
 }
To obtain \eno{pTwoPointFunction} we have assumed that $|k_i \epsilon|_q \leq 1/p$ for $i=1,2$.  Now we use \eno{fRiemannSum} to obtain
 \eqn{chiSums}{
  q \sum_{z \in S^{\Omega-1}_\mu} \chi((k_1+k_2)z) \approx
    \sum_{z \in S^\Omega_\mu} \chi((k_1+k_2)z) \approx
    q^\Omega \int_{\mathbb{Q}_q} dz \, \chi((k_1+k_2)z) = 
    q^\Omega \delta(k_1+k_2) \,,
 }
where the approximate equalities become exact in the limit where the cutoffs are removed.  Of course, we mean \eno{chiSums} in the sense that if we integrate either of the discrete sums with respect to $k_1$ against a continuous test function $\tilde{f}(k_1)$ with bounded support, the result is $\tilde{f}(-k_2)$.  Simplifying also the quantity in square brackets in \eno{pTwoPointFunction} by expanding through first order in $|\epsilon|_q^{2\Delta-n}$ (where we assume $\Delta > n/2$), we obtain
 \eqn{pTPFagain}{
  \langle {\cal O}_\epsilon(k_1) {\cal O}_\epsilon(k_2) \rangle_p = 
   {\eta_p \over |\epsilon|_p^n} \delta(k_1+k_2) \left[ -{1 \over 2\zetafct_p(2\Delta-2n)} + 
     {p^n \zeta_R \over \zetafct_p(2\Delta-n)}
     |k_1 \epsilon|_q^{2\Delta-n} + \ldots \right] \,.
 }
The omitted terms, indicated as $\ldots$, are suppressed by positive integer powers of $|k_1 \epsilon|_q^{2\Delta-n}$ relative to the last term shown.  To return from Fourier space to position space, we start by inverting the Fourier transform \eno{PFeval} and then take $z_0 \to 0$ ($p$-adically) to obtain
 \eqn{pFourierInt}{
  \zeta_R \int_{\mathbb{Q}_q} dk \, \chi(kx) |k|_q^{2\Delta-n} = 
    {\zetafct_p(2\Delta) \over \zetafct_p(2\Delta-n)} {1 \over |x|_q^{2\Delta}} \,,
 }
up to a divergent term proportional to $\delta(x)$.  Thus (for separated points) we find
 \eqn{pTPFposition}{
  \langle {\cal O}_\epsilon(x_1) {\cal O}_\epsilon(x_2) \rangle_p = 
    \eta_p |\epsilon|_p^{2\Delta-2n} {p^n \zetafct_p(2\Delta) \over \zetafct_p(2\Delta-n)^2}
     {1 \over |x_{12}|^{2\Delta}} \,.
 }
Because $K(\epsilon,k) \approx |\epsilon|_p^{n-\Delta}$ for small $\epsilon$, we introduce a leg factor in the $p$-adic case,
 \eqn{pOlocDef}{
  {\cal O}(x) = \lim_{\epsilon \to 0} |\epsilon|_p^{n-\Delta} {\cal O}_\epsilon(x) \,,
 }
and correspondingly the two-point function for the local operator ${\cal O}(x)$ is
 \eqn{pOOexclusive}{
  \langle {\cal O}(x_1) {\cal O}(x_2) \rangle_{p,\rm exclusive} = 
   \eta_p {p^n \zetafct_p(2\Delta) \over \zetafct_p(2\Delta-n)^2}
     {1 \over |x_{12}|^{2\Delta}} \,,
 }
up to contact terms.  We use the notation ``${p,\rm exclusive}$'' in \eno{pOOexclusive} as a reminder that we employed a specific cutoff procedure, namely to restrict $|\epsilon|_p < |a_0|_p$ in the outer sum of \eno{Spadic}, which excludes the points right at the boundary $|\epsilon|_p = |a_0|_p$.  Carrying through the whole computation with the restriction $|\epsilon|_p \leq |a_0|_p$ which includes these boundary points, we obtain instead
 \eqn{pOOinclusive}{
  \langle {\cal O}(x_1) {\cal O}(x_2) \rangle_{p,\rm inclusive} = 
   p^{2\Delta-2n} \langle {\cal O}(x_1) {\cal O}(x_2) \rangle_{p,\rm exclusive} \,.
 }
We see no reason to prefer the condition $|\epsilon|_p \leq |a_0|_p$ over $|\epsilon|_p < |a_0|_p$, or vice versa.  We therefore take the democratic approach of taking the geometric mean of \eno{pOOexclusive} and \eno{pOOinclusive} to get our final result:
 \eqn{pOOfinal}{
  \langle {\cal O}(x_1) {\cal O}(x_2) \rangle_p = 
   \eta_p {p^\Delta \zetafct_p(2\Delta) \over \zetafct_p(2\Delta-n)^2}
     {1 \over |x_{12}|^{2\Delta}} \,,
 }
again up to contact terms.  We will comment further on the prefactor in \eno{pOOfinal} in section~\ref{DISCUSSION}.

\subsection{Contact diagrams and higher-point correlators}

A crucial ingredient in higher-point correlation functions is contact diagrams, which in the Archimedean place are diagrammatic representations of amplitudes
 \eqn{ContactThree}{
  A_\infty(\vec{x}_1,\vec{x}_2,\vec{x}_3) \equiv
    \int {d^{n+1} y \over y_0^{n+1}} \prod_{i=1}^3 \hat{K}(y_0,\vec{y}-\vec{x}_i)
   = {\zetafct_\infty(\Delta)^3 \zetafct_\infty(3\Delta-n) \over 2 \zetafct_\infty(2\Delta)^3}
      {1 \over |\vec{x}_{12}|^\Delta |\vec{x}_{23}|^\Delta |\vec{x}_{13}|^\Delta}
 }
for three-point functions, and
 \eqn{ContactFour}{
  D_\infty(\vec{x}_1,\vec{x}_2,\vec{x}_3,\vec{x}_4) \equiv
    \int {d^{n+1} y \over y_0^{n+1}} \prod_{i=1}^4 \hat{K}(y_0,\vec{y}-\vec{x}_i)
 }
for four-point functions.  We would like to consider the analogous $p$-adic amplitudes.  We will not consider the exchange diagram for four-point functions.

\begin{figure}
  \begin{picture}(468,405)(0,-25)
   \put(30,116){\includegraphics[width=2.5in]{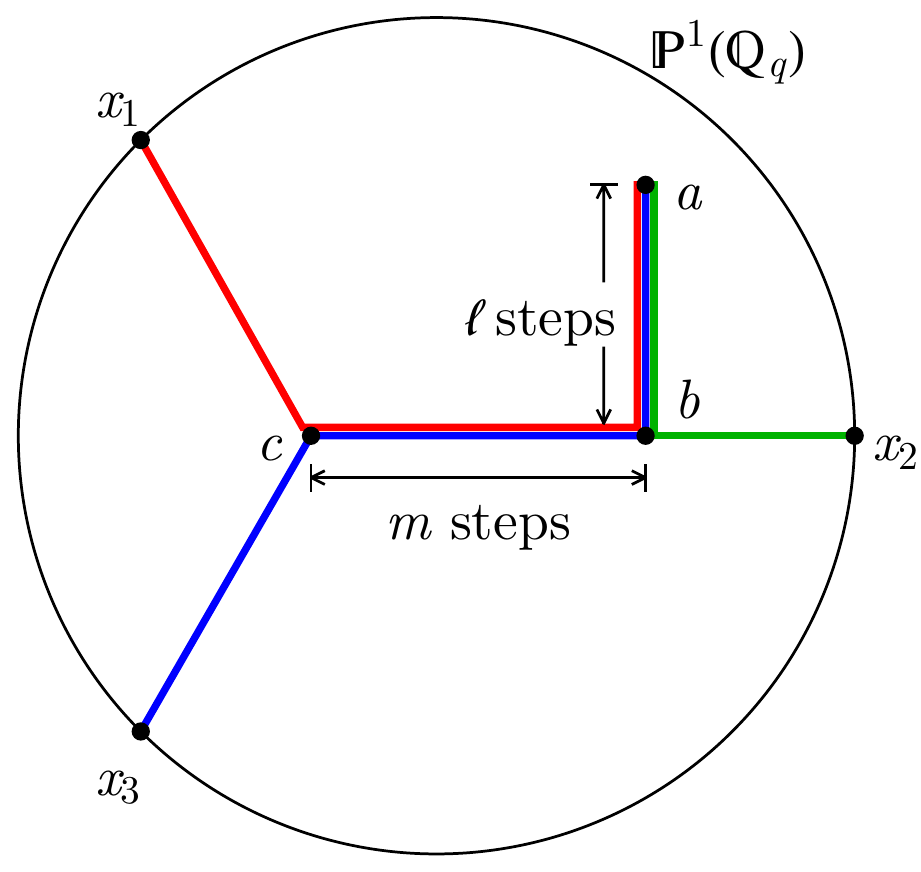}}
   \put(105,95){\Large (a)}
   \put(260,220){\includegraphics[width=2.5in]{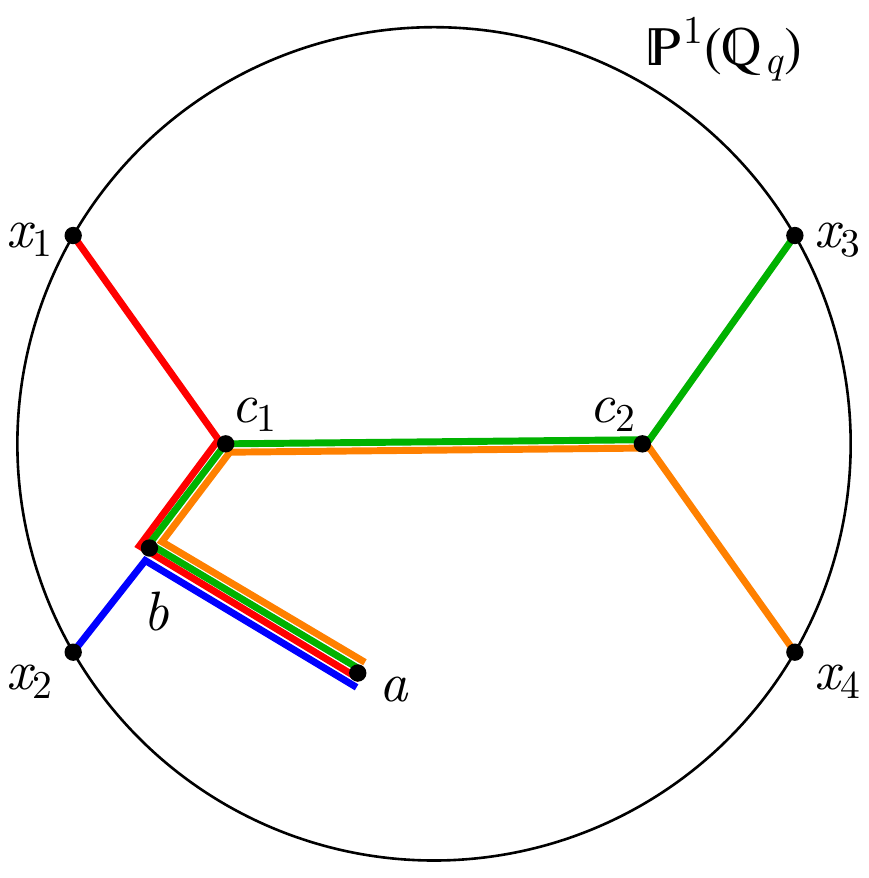}}
   \put(340,200){\Large (b)}
   \put(260,0){\includegraphics[width=2.5in]{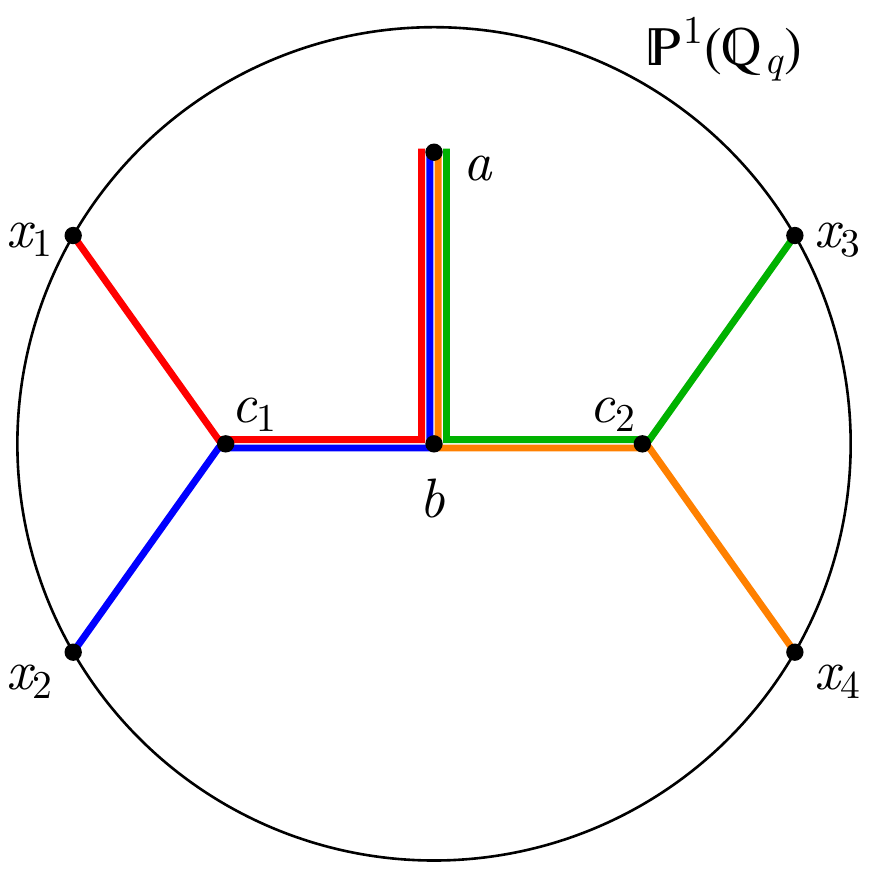}}
   \put(340,-20){\Large (c)}
  \end{picture}
  \caption{(Color online.)  Subway diagrams, indicating disjoint unions of paths on $T_q \sqcup \partial T_q$.  (a) Paths from $x_1$, $x_2$, and $x_3$ meet at the bulk point $c \in T_q$ and comprise what we refer to as the main tree.  The product $\prod_{i=1}^3 \hat{K}(a,x_i)$ relates to paths from the $x_i$ which all go to the point $a$ after first passing through the point $b$, which is the projection of $a$ onto the main tree.  (b) and (c): Paths from $x_1$ and $x_2$ to $x_3$ and $x_4$ overlap between $c_1$ and $c_2$.  There are two classes of subway diagrams contributing to the four-point amplitude, depending on whether the projection $b$ of $a$ onto the main trunk falls between $c_1$ and $c_2$ or on a leg between some $x_i$ and the appropriate $c_j$.}\label{figContactThree}
 \end{figure}
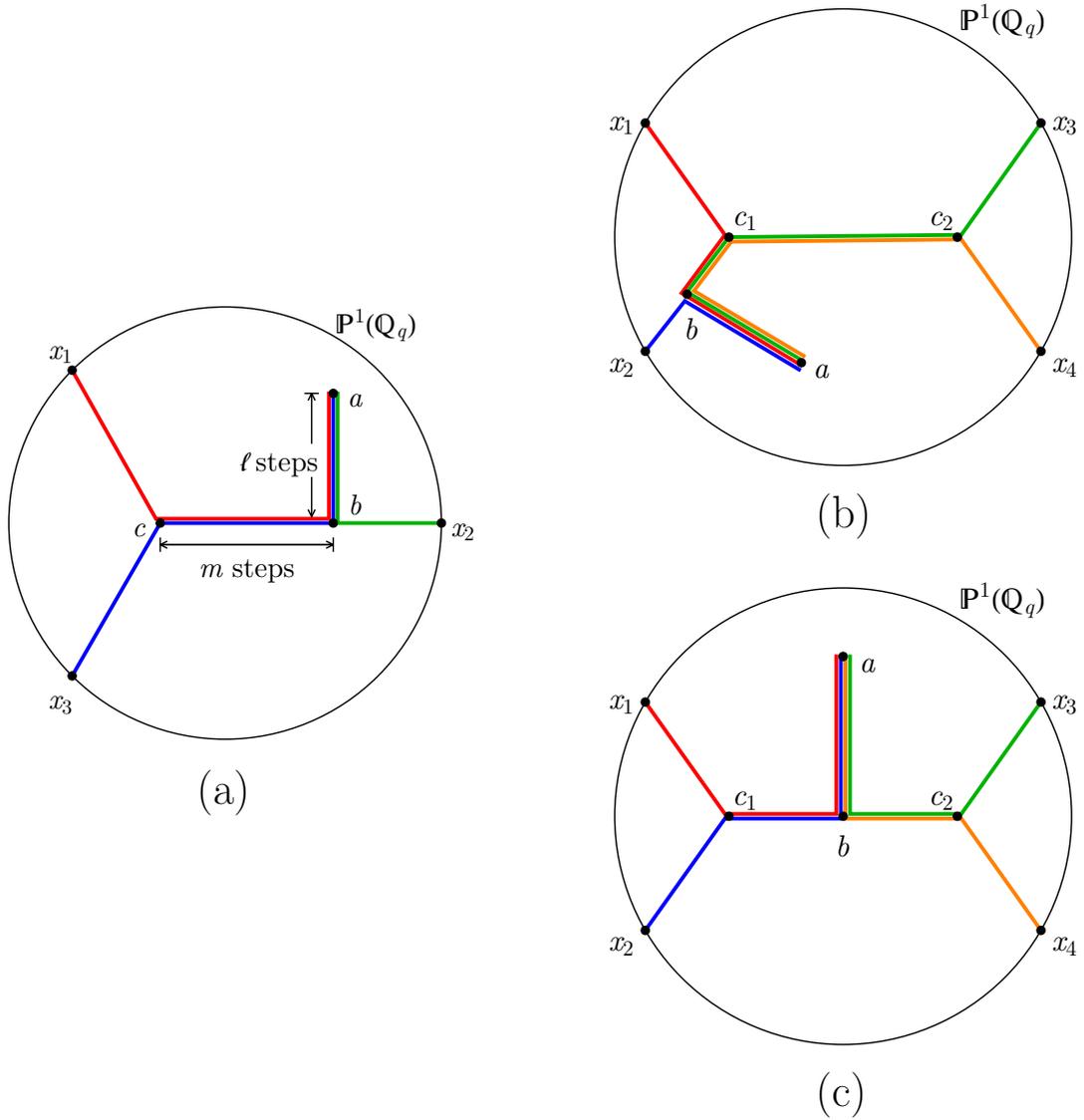

For the three-point function, the first observation is that three non-coincident points $x_1$, $x_2$, and $x_3$ in $\mathbb{P}^1(\mathbb{Q}_q)$ determine a unique point $c \in T_q$ as the point where paths from $x_1$, $x_2$, and $x_3$ meet.  We may therefore express
 \eqn{pContactThree}{
  A_p(x_1,x_2,x_3) \equiv \sum_{a \in T_q} \prod_{i=1}^3 \hat{K}(a,x_i)
   = \left[ \prod_{i=1}^3 \hat{K}(c,x_i) \right] \sum_{a \in T_q} \hat{G}(c,b) \hat{G}(b,a)^3 \,,
 }
where in every term of the sum, $b$ is the point where a path from $a$ to $c$ first joins one of the paths from the $x_i$ to $c$: see the so-called subway diagram, figure~\ref{figContactThree}a.
To demonstrate the second equality in \eno{pContactThree}, let the path (with no backtracking) in $T_q \sqcup \partial T_q$ from $x$ to $y$ be denoted $(x:y)$.  Consider a path to be a collection of edges.  Then for every $a \in T_q$, we have
 \eqn{DisjointUnion}{
  \bigsqcup_{i=1}^3 (x_i:a) = \bigsqcup_{i=1}^3 (x_i:c) \sqcup (b:c) \sqcup 3 (a:b) \,.
 }
Here we are using $\sqcup$ to form a union in which the multiplicity of each element is counted.  For example, if an edge $e$ is in $A$ with multiplicity $2$ and $B$ with multiplicity $1$, it is in $A \sqcup B$ with multiplicity $3$.  Of course, $3 (a:b)$ means $(a:b) \sqcup (a:b) \sqcup (a:b)$.  Noting that each edge leads to a factor of $p^{-\Delta}$, we arrive at \eno{pContactThree}.  Figure~\ref{figContactThree}a illustrates how this works for a particular point $a \in T_q$.  Using \eno{KGunnormalized}, we see that
 \eqn{Khelpful}{
  \hat{K}(c,x_1) = \left| {x_{23} \over x_{12} x_{13}} \right|_q^\Delta \,,
 }
with similar expressions for $\hat{K}(c,x_2)$ and $\hat{K}(c,x_3)$.  Thus, straightforwardly we find
 \eqn{xDependenceThree}{
  \prod_{i=1}^3 \hat{K}(c,x_i) = {1 \over |x_{12} x_{23} x_{13}|_q^\Delta} \,.
 }
We will refer to the union of the three paths $(x_i,c)$ as the main tree, and then $b$ can be thought of as the projection of $a$ onto the main tree.

In order to work out the sum in \eno{pContactThree}, we denote
 \eqn{ellmDefs}{
  \ell = d(a,b) \qquad\qquad m = d(b,c) \,.
 }
Then if $m=0$, meaning that $b$ and $c$ coincide, there is one point with $\ell=0$ (namely $a=c$), and there are $(p^n-2) p^{n(\ell-1)}$ points $a$ at a fixed distance $\ell>0$ from $c$ whose projection onto the main tree is $c$.  On the other hand, if $m>0$, then there are three possible choices for $b$.  Once the choice of $b$ is made, there is a single point with $\ell=0$ (namely $a=b$), and there are $(p^n-1) p^{n(\ell-1)}$ points $a$ at a fixed distance $\ell>0$ from $b$ whose projection onto the main tree is $b$.  Therefore
 \eqn{GGsum}{
  \sum_{a \in T_q} \hat{G}(c,b) \hat{G}(b,a)^3 &=
    3 \sum_{m=1}^\infty p^{-\Delta m} \left[ 1 + 
      \sum_{\ell=1}^\infty (q-1) q^{\ell-1} \left(p^{-\Delta\ell}\right)^3 \right]  \cr
   &\qquad{} +  
      \left[ 1 + \sum_{\ell=1}^\infty (q-2) q^{\ell-1}  \left(p^{-\Delta l}\right)^3 \right]  \cr
   &= {\zetafct_p(\Delta)^3 \zetafct_p(3\Delta-n) \over \zetafct_p(2\Delta)^3} \,,
 }
where for convergence we must require $\Delta > n/3$, which is certainly true since we choose the root $\Delta = \Delta_+ > n/2$.  To summarize,
 \eqn{pCTR}{
  A_p(x_1,x_2,x_3) = {\zetafct_p(\Delta)^3 \zetafct_p(3\Delta-n) \over \zetafct_p(2\Delta)^3}
    {1 \over |x_{12} x_{23} x_{13}|_q^\Delta} \,.
 }
With the three-point amplitude \eno{pCTR} in hand, we can give an account of three-point correlators of the operator ${\cal O}$ dual to $\phi$.  By the same arguments as used in the real case \cite{Muck:1998rr,Freedman:1998tz}, one finds
 \eqn{OOOcorrelator}{
  \langle {\cal O}(x_1) {\cal O}(x_2) {\cal O}(x_3) \rangle_p = -\eta_p g_3 
   {\zetafct_p(2\Delta)^3 \over \zetafct_p(2\Delta-n)^3} A_p(x_1,x_2,x_3) \,.
 }

To study the four point amplitude
 \eqn{pContactFour}{
  D_p(x_1,x_2,x_3,x_4) = \sum_{a \in T_q} \prod_{i=1}^4 \hat{K}(a,x_i) \,,
 }
let us first stipulate that $|(x_{12} x_{34}) / (x_{13} x_{24})|_q < 1$, so that the paths among the $x_i$ on $T_q$ have the topology shown in figures~\ref{figContactThree}b and~\ref{figContactThree}c: The paths from $x_1$ and $x_2$ meet at $c_1$; the paths from $x_3$ and $x_4$ meet at $c_2$; and the separation of the bulk points $c_1$ and $c_2$ is
 \eqn{cDistance}{
  d(c_1,c_2) = -\log_p \left| {x_{12} x_{34} \over x_{13} x_{24}} \right|_q = 
   -\log_p \left| {x_{12} x_{34} \over x_{14} x_{23}} \right|_q \,,
 }
where $\log_p$ is the base $p$ logarithm.  (Note that the middle expression in \eno{cDistance} is also the valuation of $(x_{12} x_{34}) / (x_{13} x_{24})$.)  Just as in \eno{pContactThree}, we may decompose the amplitude into the product of an $x_i$--dependent part based on the main tree $(x_1:c_1) \sqcup (x_2:c_1) \sqcup 2(c_1:c_2) \sqcup (x_3:c_2) \sqcup (x_4:c_2)$ times a prefactor expressed as a sum  over $T_q$:
 \eqn{pCFagain}{
  D_p(x_1,x_2,x_3,x_4) = \left[ \hat{K}(c_1,x_1) \hat{K}(c_1,x_2) \hat{K}(c_2,x_3)
    \hat{K}(c_2,x_4) \hat{G}(c_1,c_2)^2 \right] \hat{D}_p
 }
where
 \eqn{Dhat}{
  \hat{D}_p &= 4 \sum_{m=1}^\infty p^{-2\Delta m} 
    \left[ 1 + \sum_{\ell=1}^\infty (q-1) q^{\ell-1} (p^{-\Delta\ell})^4 \right]  \cr
   &\qquad{} + (d(c_1,c_2) - 1)
    \left[ 1 + \sum_{\ell=1}^\infty (q-1) q^{\ell-1} (p^{-\Delta\ell})^4 \right]  \cr
   &\qquad{} + 2
    \left[ 1 + \sum_{\ell=1}^\infty (q-2) q^{\ell-1} (p^{-\Delta\ell})^4 \right]  \cr
   &= \left[ -{1 \over \zetafct_p(4\Delta)} \log_p 
     \left| {x_{12} x_{34} \over x_{13} x_{24}} \right|_q + 
     \left( {\zetafct_p(2\Delta) \over \zetafct_p(4\Delta)} + 1 \right)^2 - 3 \right]
      \zetafct_p(4\Delta-n) \,,
 }
and we need $\Delta > n/4$ for convergence, which is always the case for $\Delta=\Delta_+$.  The first line of \eno{Dhat} comes from configurations where $b$ is on one of the legs $(x_i:c_j)$ of the main trunk, as in figure~\ref{figContactThree}b.  The second line of \eno{Dhat} comes from configurations where $b$ is on the connecting leg $(c_1:c_2)$, as in figure~\ref{figContactThree}c.  The third line comes from configurations where $b=c_1$ or $b=c_2$.  To simplify the factor in square brackets in \eno{pCFagain}, we use relations
 \eqn{KGfour}{
  \hat{K}(c_1,x_1) = \left| {x_{23} \over x_{12} x_{13}} \right|_q^\Delta = 
    \left| {x_{24} \over x_{12} x_{14}} \right|_q^\Delta \qquad\qquad
  \hat{G}(c_1,c_2) = \left| {x_{12} x_{34} \over x_{13} x_{24}} \right|_q^\Delta = 
    \left| {x_{12} x_{34} \over x_{14} x_{23}} \right|_q^\Delta \,.
 }
There are similar relations for the other factors of $\hat{K}$ in \eno{pCFagain}.  Combining them, we find
 \eqn{pCFfinal}{
  D_p(x_1,x_2,x_3,x_4) = 
   \left[ -{1 \over \zetafct_p(4\Delta)} \log_p 
     \left| {x_{12} x_{34} \over x_{13} x_{24}} \right|_q + 
     \left( {\zetafct_p(2\Delta) \over \zetafct_p(4\Delta)} + 1 \right)^2 - 3 \right]
  {\zetafct_p(4\Delta-n) \over |x_{13} x_{24}|_q^{2\Delta}} \,.
 }
We derived \eno{pCFfinal} on the assumption $|(x_{12} x_{34}) / (x_{13} x_{24})|_q < 1$.  If instead $|(x_{12} x_{34}) / (x_{13} x_{24})|_q = 1$, then all four paths from the $x_i$ meet at a common vertex $c$, and by an explicit calculation similar to \eno{GGsum}, it is straightforward to check that \eno{pCFfinal} still holds as written.  (Of course, the $\log_p$ term vanishes identically.)  Amusingly, this degeneration is impossible for $q=2$.  By relabeling the $x_i$ if necessary, we can always reach a situation where $|(x_{12} x_{34}) / (x_{13} x_{24})|_q \leq 1$, so \eno{pCFfinal} is in fact a general result.  Note that it is also an exact result, which shows that $D_p$ is considerably simpler than $D_\infty$.  However, the leading logarithmic term of $D_\infty$ for extreme values of the argument of the log essentially agrees with the $\log_p$ term in \eno{pCFfinal}, as we will now show.

The leading logarithmic part of $D_\infty$ can be extracted from the expression for $D_\infty$ in \cite{DHoker:1999pj} written as a series expansion in powers of conformally invariant variables $s$ and $t$,
\eqn{stDef}{
s \equiv {1 \over 2} {|\vec{x}_{13}|^2 |\vec{x}_{24}|^2 \over |\vec{x}_{12}|^2 |\vec{x}_{34}|^2 + |\vec{x}_{14}|^2 |\vec{x}_{23}|^2} \qquad 
t \equiv {|\vec{x}_{12}|^2 |\vec{x}_{34}|^2 - |\vec{x}_{14}|^2 |\vec{x}_{23}|^2 \over |\vec{x}_{12}|^2 |\vec{x}_{34}|^2 + |\vec{x}_{14}|^2 |\vec{x}_{23}|^2}\,.
}
Specializing to identical operators of dimension $\Delta$ in equations (A.1), (A.3) and (6.30) of \cite{DHoker:1999pj}, we obtain
\eqn{ExpandDinfty}{
 D_\infty(\vec{x}_1,\vec{x}_2,\vec{x}_3,\vec{x}_4)_{\log} &= {-(2s)^{\Delta} 2^{\Delta-2} \over |\vec{x}_{13}|^{2\Delta} |\vec{x}_{24}|^{2\Delta}}\, { \zetafct_\infty(4\Delta - n) \over \zetafct_\infty(2\Delta)^2}\, \log(1-t^2)\cr
 & \quad \times \sum_{\ell=0}^{\Delta -1} \sum_{k=0}^\infty {(-2)^{-\ell} \Gammafct(k+1)\, s^{\Delta-\ell-1} (1-2s)^{k+\ell-2\Delta+2} \over \Gammafct(\Delta-\ell)^2\, \ell! \Gammafct(k+\ell-2\Delta+3)}\, \alpha_k(t)
}
where
\eqn{alphakDef}{
 \alpha_k(t) = \sum_{\ell=0}^\infty {\Gammafct(\ell+1/2) \over \Gammafct(1/2)\, \ell!} {(1-t^2)^\ell \over 2\ell+k+1}\,,
 }
and now $\log$ indicates a natural logarithm.
It is noteworthy that if $|\vec{x}_{12}||\vec{x}_{34}| \ll |\vec{x}_{13}||\vec{x}_{24}|$, 
\eqn{schannelLimits}{
s \rightarrow {1 \over 2} \qquad t \rightarrow -1 \qquad (1-t^2) \rightarrow 4{|\vec{x}_{12}|^2 |\vec{x}_{34}|^2 \over |\vec{x}_{14}|^2 |\vec{x}_{23}|^2} \approx 4{|\vec{x}_{12}|^2 |\vec{x}_{34}|^2 \over |\vec{x}_{13}|^2 |\vec{x}_{24}|^2}\,,
}
 \eqn{}{
 \alpha_k(t) = {1 \over 1+k} + O(1-t^2)
 }
and the leading logarithmic singularity in \eno{ExpandDinfty} arises at $k=-\ell +2\Delta -2$ in the infinite sum. Then the leading order contribution from the second line of \eno{ExpandDinfty} evaluates to
 \eqn{ellSumDinfty}{
 \sum_{\ell=0}^{\Delta -1} {(-2)^{-\ell} \Gammafct(2\Delta -\ell -1)  \over 2^{\Delta-\ell-1} \Gammafct(\Delta-\ell)^2\, \ell! }\, {1 \over 2\Delta -\ell -1} = 2^{2-3\Delta} {\zetafct_\infty(2\Delta) \over \zetafct_\infty(2\Delta+1)}\,.
 }
Combining \eno{ellSumDinfty} with the first line of \eno{ExpandDinfty}, we obtain to leading logarithmic order
 \eqn{LogDinfty}{
  D_\infty(\vec{x}_1,\vec{x}_2,\vec{x}_3,\vec{x}_4)_{\log} = 
    -{\zetafct_\infty(4\Delta-n) \over \zetafct_\infty(4\Delta)}
      \left( \log {|\vec{x}_{12}||\vec{x}_{34}| \over |\vec{x}_{13}||\vec{x}_{24}|}
       \right) {1 \over |\vec{x}_{13}|^{2\Delta} |\vec{x}_{24}|^{2\Delta}} 
 }
for $|\vec{x}_{12}||\vec{x}_{34}| \ll |\vec{x}_{13}||\vec{x}_{24}|$. The corresponding expressions in the $|\vec{x}_{14}||\vec{x}_{23}| \ll |\vec{x}_{13}||\vec{x}_{24}|$ and $|\vec{x}_{13}||\vec{x}_{24}| \ll |\vec{x}_{12}||\vec{x}_{34}|$ limits can be obtained by appropriately relabeling the $\vec{x}_i$ in \eno{LogDinfty}.

If $g_3=0$ so that only the contact diagram contributes to the four-point function, then standard reasoning leads to
 \eqn{OOOOform}{
  \langle {\cal O}(x_1) {\cal O}(x_2) {\cal O}(x_3) {\cal O}(x_4) \rangle_p =
    -\eta_p g_4 {\zetafct_p(2\Delta)^4 \over \zetafct_p(2\Delta-n)^4} D_p(x_1,x_2,x_3,x_4) \,.
 }
If $g_3 \neq 0$, then there are exchange diagrams.  We expect that, analogous to the real case \cite{DHoker:1999pj}, the full four-point function can be reduced to a sum of contact diagrams, some of them generalizing the $D_p$ amplitude we have worked out explicitly.

\section{Discussion}
\label{DISCUSSION}

In $p$-adic AdS/CFT, the Bruhat--Tits tree plays the role of anti-de Sitter space, while the $p$-adic numbers replace the reals.  In the simplest case, we eschew any extension of $\mathbb{Q}_p$, and then the relation between $T_p$ and $\mathbb{Q}_p$ is like the relation between the upper half plane and the reals.  In other words, unextended $p$-adic AdS/CFT is best compared to ordinary (Euclidean) ${\rm AdS}_2/{\rm CFT}_1$.  Passing to the unramified extension $\mathbb{Q}_q$, where $q = p^n$, we have suggested that there is a natural comparison to Euclidean ${\rm AdS}_{n+1}/{\rm CFT}_n$.  The obvious point in favor of this comparison is that $\mathbb{Q}_q$ is an $n$-dimensional vector space over $\mathbb{Q}_p$ with dimension $n$ and a natural norm $|\cdot|_q$ with the property $|x|_q \geq 0$ with equality iff $x=0$. Likewise, $T_q$ can be thought of as having $n$ dimensions in the directions parallel to the boundary; more technically, the edges rising up from a given vertex of $T_q$ toward $\mathbb{Q}_q$ are enumerated by elements of $\mathbb{F}_q$, which is a vector space of dimension $n$ over $\mathbb{F}_p$.  On the other hand, the natural analog of the conformal group for $\mathbb{Q}_q$ is ${\rm PGL}(2,\mathbb{Q}_q)$, which seems closer to ${\rm SL}(2,\mathbb{R})$ than to ${\rm O}(n+1,1,\mathbb{R})$.  Thus, field theories over $\mathbb{Q}_q$ are expected to be similar to $n$-dimensional Archimedean field theories, but they may possess simplifying features comparable to low-dimensional conformal field theories.  Our main results, as summarized in section~\ref{MainResults}, reinforce these expectations.  We continue in section~\ref{Comparisons} with some comparisons between standard Archimedean results and our new $p$-adic results.  Then we discuss the geometry of chordal distance in section~\ref{Chordal}, and we give a brief account of long thin Wilson loops in section~\ref{Wilson}.  We finish with some thoughts on future directions in section~\ref{Future}.

\subsection{Main results}
\label{MainResults}

Here are our main results on propagators and correlators:
 \begin{itemize}
  \item The relationship between mass and dimension is
 \eqn{MassAndDimension}{\seqalign{\span\TL & \span\TR &\qquad\span\TT}{
  m_\infty^2 L^2 &= \Delta (\Delta-n) & for $v=\infty$  \cr
  m_p^2 &= -{1 \over \zetafct_p(\Delta-n) \zetafct_p(-\Delta)} & for $v=p$\,.
 }}
(The local zeta functions $\zetafct_\infty$ and $\zetafct_p$ were introduced in \eno{zetaInfty} and \eno{zetaForm}, respectively.)
  \item The bulk-to-boundary propagator is
 \eqn{BulkBoundarySummary}{\seqalign{\span\TL & \span\TR &\qquad\span\TT}{
  K(z_0,\vec{z};\vec{x}) &= {\zetafct_\infty(2\Delta) \over \zetafct_\infty(2\Delta-n)}
    {z_0^\Delta \over (z_0^2 + (\vec{z}-\vec{x})^2)^\Delta} & for $v=\infty$  \cr
  K(z_0,z;x) &= {\zetafct_p(2\Delta) \over \zetafct_p(2\Delta-n)}
    {|z_0|_p^\Delta \over |(z_0,z-x)|_s^{2\Delta}} & for $v=p$\,,
 }}
  \item The bulk-to-bulk propagator is
 \eqn{BulkBulkSummary}{\seqalign{\span\TL & \span\TR &\qquad\span\TT}{
  G(z_0,\vec{z};w_0,\vec{w}) &= {1 \over 2\Delta-n} {\zetafct_\infty(2\Delta) \over
    \zetafct_\infty(2\Delta-n)} u_\infty^{-\Delta}  \cr
     &\qquad{} \times 
     {}_2F_1\left( \Delta,\Delta-n+{1 \over 2};2\Delta-n+1;-{4 \over u_\infty} \right) &
     for $v=\infty$  \cr
  G(z_0,z;w_0,w) &= {\zetafct_p(2\Delta-n) \over p^\Delta} 
    {\zetafct_p(2\Delta) \over \zetafct_p(2\Delta-n)} u_p^{-\Delta} & for $v=p$\,,
 }}
where we define
 \eqn{ChordalDistance}{
  u_\infty \equiv {(z_0-w_0)^2+(\vec{z}-\vec{w})^2 \over z_0 w_0} \qquad\qquad
  u_p \equiv p^{d(z_0,z;w_0,w)} \,.
 }
  \item The two-point function is
 \eqn{TwoPointFunctions}{\seqalign{\span\TL & \span\TR &\qquad\span\TT}{
  \langle {\cal O}(\vec{x}_1) {\cal O}(\vec{x}_2) \rangle_\infty &= 
    \eta_\infty L^{n-1} (2\Delta-n) {\zetafct_\infty(2\Delta) \over \zetafct_\infty(2\Delta-n)}
      {1 \over |\vec{x}_{12}|^{2\Delta}} & for $v=\infty$  \cr
  \langle {\cal O}(x_1) {\cal O}(x_2) \rangle_p &=
    \eta_p {p^\Delta \over \zetafct_p(2\Delta-n)} 
    {\zetafct_p(2\Delta) \over \zetafct_p(2\Delta-n)} {1 \over |x_{12}|_q^{2\Delta}}
     & for $v=p$\,.
 }}
Recall that in computing the two-point function for the $p$-adics, we faced some arbitrariness in the prefactor based on the precise cutoff scheme we employed.  The result in \eno{TwoPointFunctions} is based on the democratic approach of geometrically averaging over the inclusive and exclusive cutoff scheme as explained around \eno{pOOexclusive}--\eno{pOOinclusive}.
  \item The three-point function is
 \eqn{ThreePointFunctions}{\seqalign{\span\TL & \span\TR & \qquad\span\TT}{
  \langle {\cal O}(\vec{x}_1) &{\cal O}(\vec{x}_2) {\cal O}(\vec{x}_3) \rangle_\infty  \cr
   &= -\eta_\infty L^{n-1} g_3 {\zetafct_\infty(\Delta)^3 \zetafct_\infty(3\Delta-n) \over
     2\zetafct_\infty(2\Delta-n)^3} 
       {1 \over |\vec{x}_{12}|^\Delta |\vec{x}_{23}|^\Delta |\vec{x}_{13}|^\Delta}
      & for $v=\infty$  \cr
  \langle {\cal O}(x_1) &{\cal O}(x_2) {\cal O}(x_3) \rangle_p  \cr
   &= -\eta_p g_3 {\zetafct_p(\Delta)^3 \zetafct_p(3\Delta-n) \over \zetafct_p(2\Delta-n)^3} 
       {1 \over |x_{12} x_{23} x_{13}|_q^\Delta}
      & for $v=p$\,.
 }}
  \item The four-point function is built from contact diagrams, the simplest of which have leading logarithmic singularities of the form
 \eqn{FourPointLogs}{\seqalign{\span\TL & \span\TR & \qquad\span\TT}{
  D_\infty(\vec{x}_1,&\vec{x}_2,\vec{x}_3,\vec{x}_4)_{\log}  \cr &=
   -{\zetafct_\infty(4\Delta-n) \over \zetafct_\infty(4\Delta)}
     \left( \log {|\vec{x}_{12}||\vec{x}_{34}| \over |\vec{x}_{13}||\vec{x}_{24}|} \right)
     {1 \over |\vec{x}_{13}|^{2\Delta} |\vec{x}_{24}|^{2\Delta}} & for $v=\infty$  \cr
  D_p(x_1,&x_2,x_3,x_4)_{\log}  \cr &=
   -{\zetafct_p(4\Delta-n) \over \zetafct_p(4\Delta)}
     \left( \log_p \left| {x_{12} x_{34} \over x_{13} x_{24}} \right|_q \right)
     {1 \over |x_{13} x_{24}|_q^{2\Delta}} & for $v=p$\,.
 }}
 \end{itemize}
The Archimedean results in \eno{MassAndDimension}--\eno{FourPointLogs} are standard in the literature, although their simplified presentation in terms of the local zeta function $\zetafct_\infty(s)$ is new as far as we are aware.

There are some natural simplifying features of $p$-adic AdS/CFT.  To begin with, all quantities of interest lie in the field extension $\mathbb{Q}(p^\Delta)$: that is, rational numbers combined with integer powers of $p^\Delta$.  Many of the powers of $p$ can be efficiently packaged in the $p$-adic zeta function $\zetafct_p$, and Archimedean results can be (almost) recovered by replacing $p$ by $\infty$.  The reason why the $p$-adic formulas we derived are valued in $\mathbb{Q}(p^\Delta)$ is that they come from sums of products of propagators over the Bruhat--Tits tree, and these sums typically reduce to geometric series, $\sum_{\ell=1}^\infty x^\ell = {x \over 1-x}$, with $x \in \mathbb{Q}(p^\Delta)$, and the map $x \to {x \over 1-x}$ involves only field operations.  It would be interesting to see whether more sophisticated computations in $p$-adic AdS/CFT lead to results valued outside $\mathbb{Q}(p^\Delta)$.  In the Archimedean place, we are struck by the complete absence of factors of $\pi$ when we express correlators in terms of $\zetafct_\infty$ rather than $\Gammafct$.  This is analogous to seeing $p$-adic correlators taking values in $\mathbb{Q}(p^\Delta)$.

\subsection{Comparing Archimedean and $p$-adic results}
\label{Comparisons}

It is clear from \eno{MassAndDimension}--\eno{FourPointLogs} that we generally cannot write closed-form expressions for physical quantities which are valid equally for $v=\infty$ and $v=p$.  We are often close to being able to do so, as in the cases of the bulk-to-boundary propagator \eno{BulkBoundarySummary}, the three-point function \eno{ThreePointFunctions}, and the leading-log term in the four-point amplitude \eno{FourPointLogs}; note however the mismatched factor of $2$ in \eno{ThreePointFunctions}, which is operationally related to the fact that we integrate over $y_0>0$ in the Archimedean calculations, whereas the sum over $T_q$ can be thought of as including an integral over all non-zero $p$-adic numbers $y_0$ (more on this later).

To better understand the relation between $v=\infty$ and $v=p$, consider first the growth of volume in Euclidean ${\rm AdS}_{n+1}$ with radius as compared to the growth in the number of vertices of $T_q$ with radius.  The volume of a ball ${\cal B}_{n+1}(R)$ of radius $R$ in ${\rm AdS}_{n+1}$ is
\eqn{volH}{
 {\rm vol}({\cal B}_{n+1}(R)) \sim {\rm constant} \times {\rm e}^{n R/L} 
   \qquad\hbox{for $R \gg L$} \,,
}
where $L$ is the radius of curvature of ${\rm AdS}_{n+1}$.  On the other hand, introducing a lattice spacing $a$ on the Bruhat--Tits tree, the volume of a ball ${\cal B}_{T_q}(R)$ of radius $R$ in $T_q$---meaning all points within a graph distance $R/a$ of a specified point---is given by
\eqn{volTq}{
{\rm vol}({\cal B}_{T_q}(R)) \sim {\rm constant} \times p^{n R/a} = 
 {\rm constant} \times {\rm e}^{n R/L_p} \qquad\hbox{for $R \gg a$\,,}
}
where we have introduced a length scale
 \eqn{LpDefine}{
  L_p \equiv {a \over \log p} \,,
 }
which stands in place of the radius of curvature $L$ and makes \eno{volH} and \eno{volTq} directly comparable.  It is helpful to see in \eno{volTq}--\eno{LpDefine} how dimensions work, but elsewhere we set $a=1$.  (Changing this to $a=1/e$ where $e$ is the ramification index could be helpful in discussing ramified extensions.)

With the length scale $L_p$ in hand, we can better understand the apparent mismatch in the two-point functions \eno{TwoPointFunctions} between the boundary factor $2\Delta-n$ in the Archimedean place and $p^\Delta/\zetafct_p(2\Delta-n)$ for the $p$-adics.  If we set $p = {\rm e}^{\log p}$ and formally treat $\log p$ as small, as in \cite{Gerasimov:2000zp,Ghoshal:2006zh}, then the $p$-adic results become, to leading order in $\log p$,
 \eqn{SmallEpsilon}{
  m_p^2 L_p^2 \approx \Delta (\Delta-n) \qquad
   {p^\Delta L_p \over \zetafct_p(2\Delta-n)} \approx 2\Delta-n \,.
 }
An improved discussion along these lines is probably possible if we extend $\mathbb{Q}_q$ by including $p^{1/e}$ and using the uniformizer $\pi = p^{1/e} = {\rm e}^{(\log p)/e}$.  In addition to \eno{SmallEpsilon}, we have
 \eqn{dmSquared}{\seqalign{\span\TL & \span\TR &\qquad\span\TT}{
  {d(m_\infty^2 L^2) \over d\Delta} &= 2\Delta-n & for $v = \infty$  \cr
  {d(m_p^2 L_p^2) \over d\Delta} &= {p^\Delta L_p \over \zetafct_p(2\Delta-n)} & for $v=p$\,,
 }}
where we have not made any formal expansion in small $\log p$.  Finally, we note that the same factor of $p^\Delta/\zetafct_p(2\Delta-n) = 2\nu_p$ also appears in the normalizations of the $p$-adic two-point function and bulk-to-bulk propagator, and in this latter context there is no cutoff-related ambiguity: See the discussion ending in \eno{nupForm}.

In short, the results of $p$-adic and Archimedean calculations have strong affinities, but we are not generally in a position to write down adelic products.  Perhaps we should not be too surprised by the mismatches between $p$-adic and Archimedean calculations, since our starting point on the $p$-adic side was only the simplest lattice action.  It seems possible that a more informed treatment of the bulk action will lead to progress toward an adelic version of AdS/CFT.

\subsection{The geometry of chordal distance}
\label{Chordal}

We believe that a good first step toward adelic AdS/CFT is to re-examine the geometry of $T_q$ from a point of view that makes its similarities to ordinary ${\rm AdS}_{n+1}$ more transparent.  Indeed, the Bruhat--Tits tree $T_q$ is a natural bulk construction both from the perspective of the representation of $p$-adic numbers as a string of digits, and from the more geometric point of view of a coset construction ${\rm PGL}(2,\mathbb{Q}_q)/{\rm PGL}(2,\mathbb{Z}_q)$, where the denominator is the maximal compact subgroup of the numerator.  But from the point of view of classical AdS/CFT, it is a bit surprising, especially since the bulk $T_q$ has a smaller cardinality than the boundary $\mathbb{Q}_q$.  Our parametrization of $T_q$ in terms of $(z_0,z)$, where $z_0 = p^\omega$ and $z \in \mathbb{Q}_q$ is known up to $O(z_0)$ corrections, suggests that it might be more natural to let the bulk be all of 
 \eqn{pAdSDef}{
  p{\rm AdS}_{n+1} \equiv \mathbb{Q}_p^\times \times \mathbb{Q}_q \,,
 }
with coordinates $(z_0,z)$ where now $z_0 \in \mathbb{Q}_p^\times$ (a non-zero $p$-adic number) and $z \in \mathbb{Q}_q$.  Introduce a chordal distance function between any two points $(w_0,w)$ and $(z_0,z)$:\footnote{The Archimedean quantity $u_\infty$ introduced in \eno{ChordalDistance} is actually the square of the chordal distance divided by $L^2$.  To be precise, if we define global coordinates
 \eqn{GlobalReal}{
  Z_0 = {1 \over 2z_0} (L^2+z_0^2 + \vec{z}^2) \qquad
  \vec{Z} = L {\vec{z} \over z_0} \qquad
  Z_{n+1} =  {1 \over 2z_0} (-L^2+z_0^2 + \vec{z}^2) \,,
 }
and similarly for $W_M$, then Euclidean ${\rm AdS}_{n+1}$ is the locus $\eta^{MN} Z_M Z_N = -L^2$, where $\eta^{MN} = \diag\{ -1,1,1,\ldots,1 \}$, and $u_\infty = \eta^{MN} (Z_M-W_M)(Z_N-W_N)/L^2$.}
 \eqn{pChordal}{
  u_p = {|(z_0-w_0,z-w)|_s^2 \over |z_0 w_0|_p} \,,
 }
where $|\cdot|_s$ indicates the supremum norm \eno{supNorm}.  It is easy to show that if $(z_0,z)$ and $(w_0,w)$ parametrize different points on $T_q$ in the sense explained in section \ref{BTTREE}, with $z_0$ and $w_0$ restricted to integer powers of $p$, then $u_p = p^{d(z_0,z;w_0,w)}$, in agreement with \eno{ChordalDistance}.  On the other hand, $u_p$ can be an arbitrarily small power of $p$ if $(w_0,w)$ and $(z_0,z)$ are $p$-adically very close to one another.  So we can ask, how is $T_q$ related to the larger space $p{\rm AdS}_{n+1}$?

It turns out there is a simple and pleasing answer: To get $T_q$, we must coarse-grain $p{\rm AdS}_{n+1}$ at the AdS scale.  Specifically, we can form an equivalence relation
 \eqn{zwEquivalence}{
  (z_0,z) \sim (w_0,w) \qquad\hbox{iff}\qquad
   u_p(z_0,z;w_0,w) \leq 1 \,.
 }
To find the equivalence classes under the relation \eno{zwEquivalence}, note that
 \eqn{xiIneq}{
   \left| \left( {z_0 \over w_0} + {w_0 \over z_0} - 2, {(z-w)^2 \over z_0 w_0} \right) \right|_s 
     = u_p \leq 1 \,,
 }
which implies that ${z_0 \over w_0} + {w_0 \over z_0} - 2$ is a $p$-adic integer.  It follows that ${w_0 \over z_0} \in \mathbb{U}_p$.  Next one can show using \eno{xiIneq} that ${z-w \over z_0}$ is also a $p$-adic integer, so $z-w \in z_0 \mathbb{Z}_q$.  In other words, for fixed $(z_0,z)$, the set of all $(w_0,w)$ with $u_p(z_0,z;w_0,w) \leq 1$ is
 \eqn{Blocks}{
  B(z_0,z) = z_0 \mathbb{U}_p \times (z + z_0 \mathbb{Z}_q) \,.
 }
In the natural measure on $p{\rm AdS}_{n+1}$, the volume of each block is the same:
 \eqn{BlockVolume}{
  \int_{B(z_0,z)} {dw_0 \, dw \over |w_0|_p^{n+1}} = 
   {1 \over |z_0|^{n+1}} \left( \int_{z_0 \mathbb{U}_p} dw_0 \right) 
    \left( \int_{z_0 \mathbb{Z}_q} dw \right) = {1 \over \zetafct_p(1)} \,,
 }
where we have used the fact that $\mathbb{U}_p$ has measure $1 - 1/p = 1/\zetafct_p(1)$.

We can label the blocks $B(z_0,z)$ uniquely by requiring $z_0 = p^\omega$ for some $\omega \in \mathbb{Z}$ and $z \in S_{-\infty}^\omega \equiv \bigcup_{\mu < \omega} S^\omega_\mu$.  The blocks $B(z_0,z)$ can now be regarded as the nodes of the tree $T_q$, and the distance function on $T_q$ is defined by
 \eqn{pdDef}{
  p^{d(z_0,z;w_0,w)} = u_p(z_0,z;w_0,w) \qquad\hbox{provided $B(z_0,z) \neq B(w_0,w)$\,,}
 }
together with the trivial definition $d(z_0,z;z_0,z) = 0$.  In the Archimedean place, the relation $u_\infty(z_0,\vec{z};w_0,\vec{w}) \leq 1$ means that $(z_0,\vec{z})$ and $(w_0,\vec{w})$ are essentially within an AdS radius of one another; thus \eno{zwEquivalence} can be regarded as a $p$-adic analog of coarse-graining at the AdS curvature scale.  However, there is no analog of the sets $B(z_0,z)$ in the Archimedean place, essentially because if we carried \eno{zwEquivalence} over to the reals, the transitive property would fail.  Less formally, we can't carve ordinary Euclidean AdS into blocks without points near the edges being very close to one another.

With the blocks $B(z_0,z)$ specified, we can go further and define a coarse topology on $p{\rm AdS}_{n+1}$ by saying that the closed sets are arbitrary unions of blocks.  Continuous functions with respect to this topology are precisely the ones which are constant on each block, which is to say well-defined as functions on $T_q$.  If $f$ is such a function, then we can calculate its integral as a sum over $T_q$:
 \eqn{fInt}{
  \zetafct_p(1) \int_{p{\rm AdS}_{n+1}} {dz_0 \, dz \over |z_0|_p^{n+1}} 
    f(z_0,z) = \sum_{a \in T_q} f(a) \,,
 }
where we have used \eno{BlockVolume}.

It is natural to inquire whether we can coarse-grain $p{\rm AdS}_{n+1}$ differently.  It is easy to see that if we try to form an equivalence relation by saying $(z_0,z) \sim (w_0,w)$ iff $u_p(z_0,z;w_0,w) \leq p^m$ for a positive integer $m$, then the transitive property fails, so we do not have a natural way to split $p{\rm AdS}_{n+1}$ into larger blocks than the ones defined in \eno{Blocks}.  On the other hand, for $m \in \mathbb{N} \equiv \{0,1,2,3,\ldots\}$, we may define
 \eqn{FinerEquivalence}{
  (z_0,z) \sim_m (w_0,w) \qquad\hbox{iff}\qquad
   u_p(z_0,z;w_0,w) \leq p^{-2m} \,,
 }
and then $\sim_m$ is an equivalence relation, and it coincides with $\sim$ when $m=0$.  (There is no point in considering $u_p(z_0,z;w_0,w) \leq p^{-2m-1}$, because $u_p$, if non-zero, must take the form $p^\sigma$ where either $\sigma \in \mathbb{N}$ or $\sigma = -2m$ for $m \in \mathbb{N}$.)  The equivalence classes $B(z_0,z)$ under $\sim$ are subdivided into smaller equivalence classes $B_m(z_0,z)$ under $\sim_m$: In other words, $\sim_m$ for $m>0$ is a refinement of $\sim$.  From the point of view of quantum gravity, such refinements are appealing because they allow us to use ``ordinary'' geometry down to a scale that we identify as the Planck scale, and length scales smaller than the Planck scale either don't exist or are qualitatively different.  An interesting point of comparison is that tensor networks in AdS based on MERA generally cannot be made finer than the AdS scale \cite{Bao:2015uaa}.

 \begin{figure}
  \begin{picture}(468,360)(0,0)
  \put(90,240){\includegraphics[width=4in]{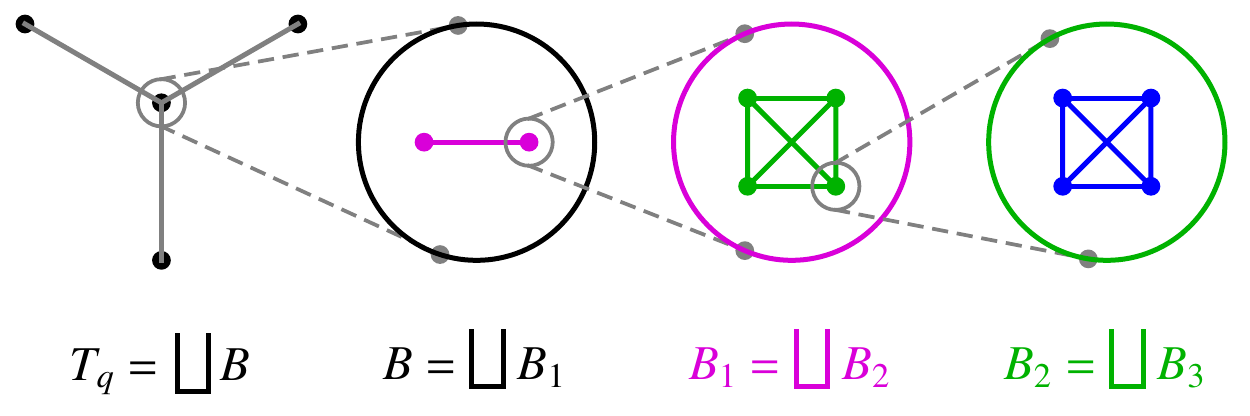}}
  \put(20,290){\Large (a)}
  \put(110,10){\includegraphics[width=3.5in]{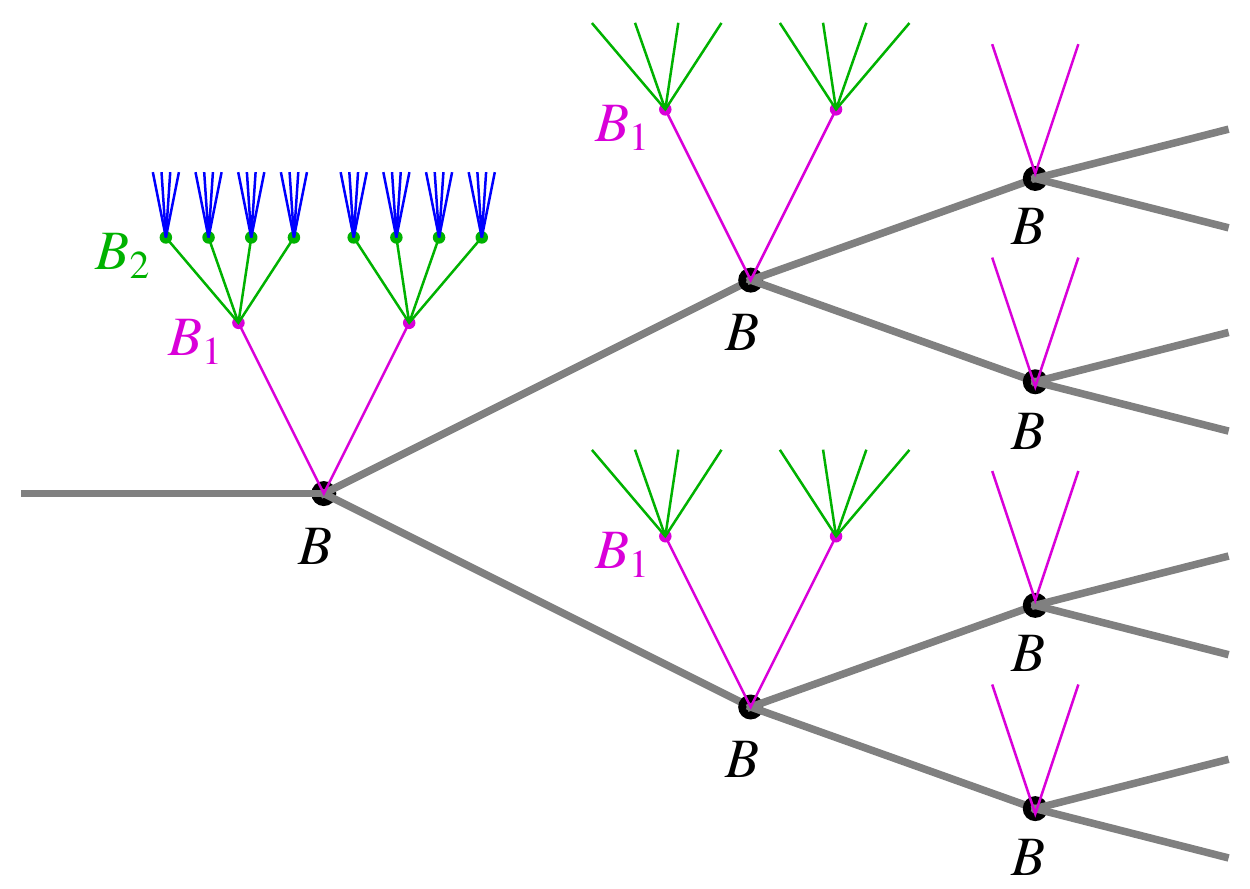}}
   \put(20,85){\Large (b)}
  \end{picture}
  \caption{(Color online.) Successive refinements of the Bruhat--Tits tree.  The example shown is for $q=p=2$.  (a) Successive refinements reveal more and more structure as we zoom in on any given bulk region.  The first step is to write $T_q$ as a disjoint union of blocks $B$ as defined in \eno{Blocks}.  The next step is to write each block $B$ as a disjoint union of blocks $B_1$ as defined in \eno{BClasses}; then each block $B_1$ is written as a disjoint union of blocks $B_2$, and so forth.  (b) Successive refinements of $T_q$ lead to the enhanced tree $T_{qp}$, in which each vertex is a block $B_m$.  The base tree $T_q$ is shown in gray, and the height $h$ measures how many steps away a point on $T_{qp}$ is from the base tree.}\label{ThreeRefinements}
 \end{figure}

To better understand the refinements described in the previous paragraph, let's examine how the block $B(1,0)$ splits into smaller blocks $B_m(z_0,z)$.  The first task is to parse the relationship $u_p(z_0,z;w_0,w) \leq p^{-2m}$ for points $(z_0,z)$ and $(w_0,w)$ in $B(1,0)$.  We have in particular $z_0,w_0 \in \mathbb{U}_p$, so
 \eqn{upSimpler}{
  u_p = |(z_0-w_0,z-w)|_s^2 \leq p^{-2m} \,,
 }
The result \eno{upSimpler} indicates that from the point of view of chordal distance, $B(1,0)$ is like a patch of $\mathbb{Q}_q \times \mathbb{Q}_p$ equipped with the supremum norm; in other words, we don't see any sign of the curvature of $p{\rm AdS}_{n+1}$ once we look at length scales within a given block $B(z_0,z)$.  From \eno{upSimpler} we see immediately that the equivalence classes of $\sim_m$ inside $B(1,0)$ are
 \eqn{BClasses}{
  B_m(z_0,z) = (z_0 + p^m \mathbb{Z}_p) \times (z + p^m \mathbb{Z}_q) \,.
 }
These equivalence classes are uniquely labeled by $(z_0,z)$ if we require 
 \eqn{zzRequire}{
  z_0 \in s^m_0 \;\backslash\; s^m_1 \qquad\hbox{and}\qquad z \in S^m_0 \,.
 }
The sets $S^\omega_\mu$ were defined in \eno{SmODef}, and by $s^\omega_\mu$ we mean the analogous sets for $\mathbb{Q}_p$ instead of $\mathbb{Q}_q$.  For $m>0$, there are $p^{m(n+1)} (1-1/p)$ equivalence classes $B_m(z_0,z)$ inside $B(1,0)$, and the volume of each one is $\int_{B_m(z_0,z)} {dw_0 dw \over |w_0|_p^{n+1}} = p^{-m(n+1)}$.  If $m=1$, then \eno{zzRequire} simplifies to $z_0 \in \mathbb{F}_p^\times = \{1,2,\ldots,p-1\}$ and $z \in \mathbb{F}_q$.  In this case, it is easy to see that $u_p = 1$ between any two points in distinct blocks $B_1(z_0,z)$ and $B_1(w_0,w)$, whereas (by definition) $u_p \leq p^{-2}$ between any two points in the same block.  If we proceed next to $m=2$, then each block $B_1(z_0,z)$ splits into $p^{n+1}$ smaller blocks $B_2(w_0,w)$, and each pair of $B_2$ blocks within a given $B_1$ block is separated by a distance $u_p = p^{-2}$.  A cartoon of these successive refinements is shown in figure~\ref{ThreeRefinements}a.  Evidently, a sequence of topologies on $p{\rm AdS}_{n+1}$ can be defined, such that functions which are constant over all the blocks $B_m(z_0,z)$ are continuous with respect to the $m^{\rm th}$ topology.  Absent a Planck scale cutoff, the endpoint of the refinement process is the full geometry $p{\rm AdS}_{n+1}$.  A simple way to think of a Planck scale cutoff is to stop refining after a finite number of steps.

The successive refinements of $T_q$ into $p{\rm AdS}_{n+1}$ can be summarized by an enhanced tree structure, constructed as follows.  From each node of the base tree $T_q$, we add $q(p-1)$ edges to indicate the splitting of $B(z_0,z)$ into blocks $B_1$.  From the terminus of each of these edges, we add $qp$ new edges to indicate the splitting of $B_1$ blocks into $B_2$ blocks.  Continuing in this way, we wind up with a tree with uniform coordination number $qp+1$, which is to say $T_{qp}$: see figure~\ref{ThreeRefinements}b.  We can identify each vertex on the enhanced tree with a block $B_m$, where the index $m$ increases the further off of the base tree we go.  Let the standard graph theoretic distance between two points $a$ and $b$ on $T_{qp}$ be denoted $D(a,b)$, and let the distance from a point $a$ on $T_{qp}$ to the nearest point on the base tree $T_q$ be denoted $h(a)$.  Then if we define
 \eqn{sigmaDef}{
  \sigma(a,b) = D(a,b) - h(a) - h(b) \,,
 }
it can be checked that the maximum value of the chordal distance $u_p$ between a point in the block associated with $a$ and a point in the block associated with $b$ is $p^{\sigma(a,b)}$.

We would like to use $p{\rm AdS}_{n+1}$ and the geometry of chordal distance as jumping off points for the construction of bulk models that are more interesting than just a scalar with nearest neighbor interactions on $T_q$ as in \eno{Snonlinear}.  Ideally we would like to have some notion of fluctuating bulk geometry.  The absence of cycles in $T_q$ makes it hard to see how to study gauge fields or Riemannian curvature.  So it is interesting to observe that at each stage of refinement, the newly introduced blocks (for example, all the $B_3$ blocks inside a given $B_2$ block) form a complete graph in the sense that each is equidistant from all the others using the chordal distance function $u_p$.  Can we take advantage of the cycles in these complete graphs to formulate some useful lattice notions of curvature?  If we can, how does curvature fit in with the structure of the enhanced tree $T_{qp}$?

\subsection{Wilson loops}
\label{Wilson}

Wilson loops are important observables in field theory. A natural question to ask is: What are the properties of $p$-adic Wilson loops? We give some preliminary indications in this section.

We begin by reviewing a simple case in the Archimedean place \cite{Maldacena:1998im, Rey:1998ik}. The interquark potential energy $V(R)$ of a heavy quark-antiquark pair is given by the expectation value of the Wilson loop operator,
 \eqn{WilsonInterquarkV}{
  \langle W(\mathcal{C}) \rangle \propto {\rm e}^{-V(R)\:T} \,,
 }
where $\mathcal{C}$ is a long thin rectangular contour, of length $T$ along the time direction, and length $R \ll T$ along the spatial direction.  The calculations are all done in Euclidean signature, so the time direction is picked out essentially arbitrarily.  According to the AdS/CFT prescription, the expectation value of a Wilson loop is given by the partition function for a string in AdS, whose edge at the boundary lies along $\mathcal{C}$. In particular, in the supergravity limit, 
 \eqn{WilsonSaddle}{
 \langle W(\mathcal{C}) \rangle \propto {\rm e}^{-(S_\Phi - \ell \Phi)}
 }
 where $S_\Phi$ is the action of the minimal surface in AdS of the string worldsheet ending on $\mathcal{C}$.  The minimal action $S_\Phi$ and hence $V(R)$ suffer from UV divergences. We must subtract away from $V(R)$ the infinite energies of the free quark and the free antiquark to get a sensible finite answer.  In other words, we must renormalize the minimal action $S_\Phi$ by subtracting from it the action of the worldsheets associated with the free quark and the free antiquark, both of which stretch all the way to infinity in AdS. This is what is meant by $\ell \Phi$ in \eno{WilsonSaddle}.  In conformal field theories, symmetry dictates that the potential energy take the form $V(R) \propto 1 / R$. Indeed, an AdS calculation of the (regulated) minimal surface yields, via \eno{WilsonInterquarkV} and \eno{WilsonSaddle}
 \eqn{VArchimedean}{
 V(R)\: T = - {4\pi^2 \over \Gammafct(1/4)^4 } {L^2 \over \alpha'} {T \over R}\,,
 }
where $1/2\pi\alpha'$ is the string tension.  We include only the $T$-extensive contribution and ignore endpoint effects having to do with how we close the loop off at times $t = \pm T/2$.  In \eno{VArchimedean}, $T$ serves as an infrared regulator imposed to avoid a divergent factor from integrating in the $t$ direction.

Let's move on to discuss a simple analog of the long thin rectangular Wilson loop in the context of $p$-adic AdS/CFT.  We restrict ourselves to a two-dimensional subspace of $\mathbb{Q}_{p^n}$, which itself is an $n$-dimensional vector space over $\mathbb{Q}_p$.  Every point $z$ in this two-dimensional subspace can be written as $z= t\: r^\ell + x\: r^k$ for fixed $\ell, k \in \{0,1,\ldots\,n-1\}$, $\ell \neq k$, and $t, x \in \mathbb{Q}_p$.  Here $r$ is a primitive $(p^n - 1)$-th root of unity. For the rest of this discussion we set $\ell=0$ and $k=1$ without loss of generality. Let the time direction be along the $r^0$ component, and let space be along $r^1$.

The $p$-adic analog of parallel lines in the $t$ direction with spatial separation $R$ is clear enough: each line is an affine map of $\mathbb{Q}_p$ to $\mathbb{Q}_q$, so that the quark line is all points of the form $t + x\: r$ with $x$ fixed and $t$ varying across $\mathbb{Q}_p$, while the antiquark line is $t + \tilde{x}\: r$, again with $\tilde{x}$ fixed and $t$ varying across $\mathbb{Q}_p$.  We anticipate the need for an infrared regulator, so we restrict the $p$-adic norm of the time coordinate: $|t|_p \leq |T|_p$,
where we require
 \eqn{padicLongThin}{
  |T|_p \gg |R|_p
 }
and $R = x-\tilde{x}$.  For convenience we set
 \eqn{TRvaluations}{
  v_p(T) = \tau \qquad\hbox{and}\qquad v_p(R) = \rho \,.
 }
Our results depend on $T$ and $R$ only through their norms, so we could set $T = p^\tau$ and $R = p^\rho$ without loss of generality.  None of our calculations below depend on this choice.

Now we must specify what we mean in the discrete context of the Bruhat--Tits tree $T_q$ by a string worldsheet whose edge is on one of the parallel lines.  Consider the quark line for specificity.  Any point $t + x\: r$ on the line is associated to a unique path through $T_q$ from $\infty$ to $t + x\: r$.  Adapting previous notation, let's denote this path as $(\infty:t + x\: r)$, with the convention that we exclude the endpoints at the boundary.  The union of these paths is a subtree of $T_q$ isomorphic to $T_p$.  This subtree is what we want to regard as the string worldsheet.  It is like a (Euclidean) $AdS_2$ subset of $AdS_{n+1}$.  To bring in our infrared regulator, let's first assume that $|T|_p>|x|_p$.  Then each path on $T_q$ from $\infty$ to a point $t + x\: r$ with $|t|_p \leq |T|_p$ passes through the point $(p^{\tau},0)$ on the main trunk of $T_q$.  We think of the infrared regulated path $((p^\tau,0):t + x \: r)$ as only that portion of the path starting at $(p^\tau,0)$ and continuing upward to $t + x\: r$.  We include $(p^\tau,0)$, but not the boundary point $t + x \: r$, in the regulated path $((p^\tau,0):t + x \: r)$.  The infrared regulated worldsheet is the union 
 \eqn{WqReg}{
  M_x(T) \equiv \bigcup_{|t|_p \leq |T|_p} ((p^\tau,0):t + x\: r) \,.
 }
The number of vertices of $T_q$ in $M_x(T)$ is still infinite, but if we discard points sufficiently close to the boundary (i.e.~impose an ultraviolet regulator) it becomes finite.

Next we need to describe in the context of $T_q$ the string worldsheet with an edge on each of the parallel lines.  To begin with, consider the $x$-direction only, and correspondingly the $T_p$ subtree at $t=0$.  The common ancestor of $x$ and $\tilde{x}$ on $T_p$ is at bulk depth $z_0 = 1/|R|_p$.  Moreover, there is a unique path leading from $x$ to $\tilde{x}$ along $T_p$, and it goes through their common ancestor.  Returning to $T_q$, we consider the string worldsheet with an edge on each of the parallel lines to be the union over $t$ of all paths in $T_q$ from $t + x\: r$ to $t + \tilde{x}\: r$:
 \eqn{CurvedWorldsheet}{
  M_{q\bar{q}}(T) \equiv \bigcup_{|t|_p \leq |T|_p} (t + x\: r: t + \tilde{x}\: r) \,,
 }
where as before we exclude boundary points from the paths.  Note that it is important that we consider only paths from $t + x\: r$ to $t + \tilde{x}\: r$ with the same value of $t$: If we allowed paths from $t + x\: r$ to $\tilde{t} + \tilde{x}\: r$ with $|t-\tilde{t}|_p > |R|_p$, then we would include points that go lower in $T_q$ than the bulk depth $p^\rho = 1/|R|_p$.  We caution that our prescription for forming the string worldsheet as a union of paths does not directly refer to minimal surfaces.  Intuitively, the worldsheet \eno{CurvedWorldsheet} is the only discretized surface with no back-tracking with edges on the quark and antiquark lines.  Closer consideration of how to describe more general string worldsheets in $T_q$ is clearly merited.
  \begin{figure}
   \begin{picture}(468,500)(0,0)
 \put(-25,330){\includegraphics[width=3.5in]{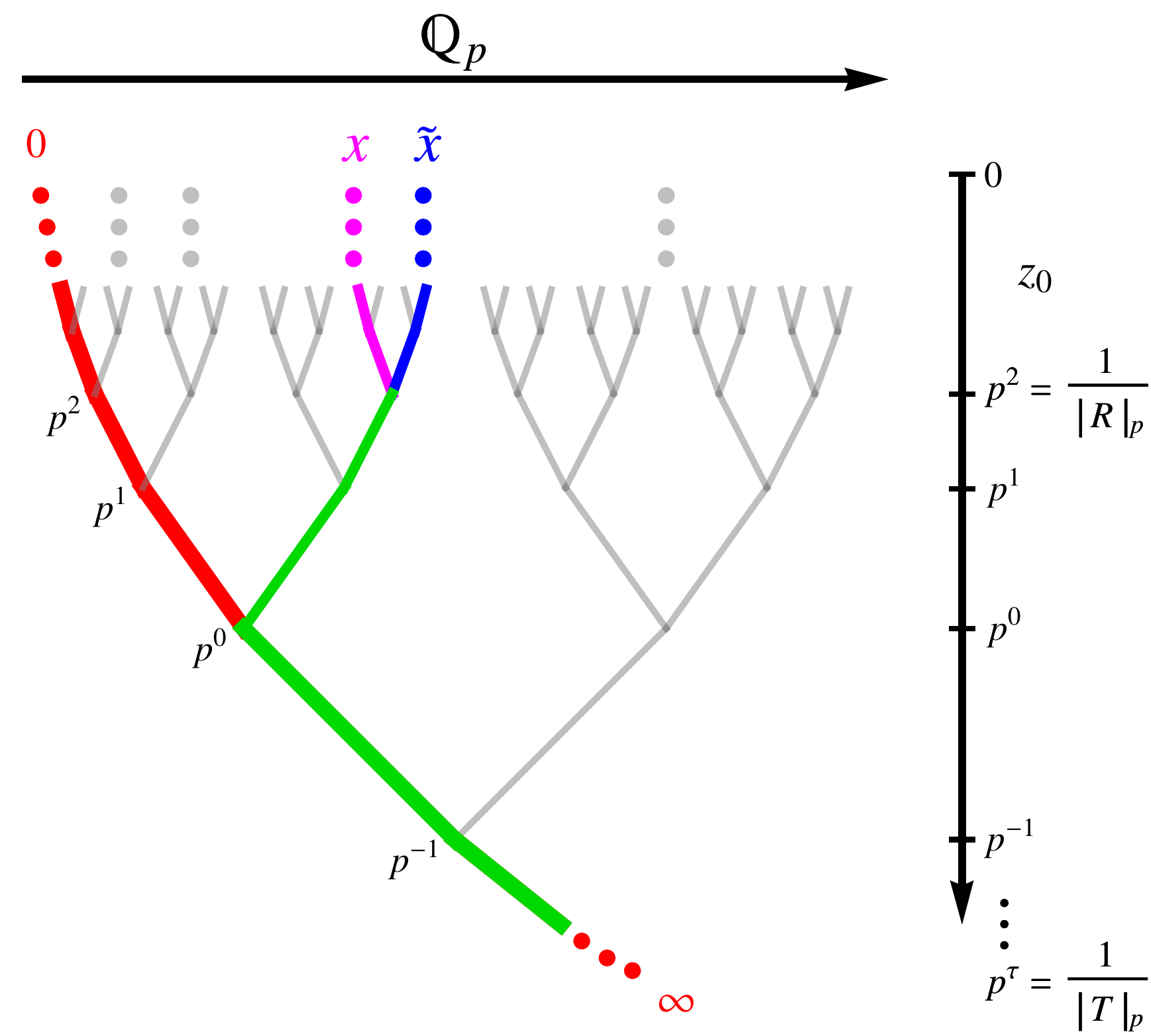}}
  \put(240,330){\includegraphics[width=3.5in]{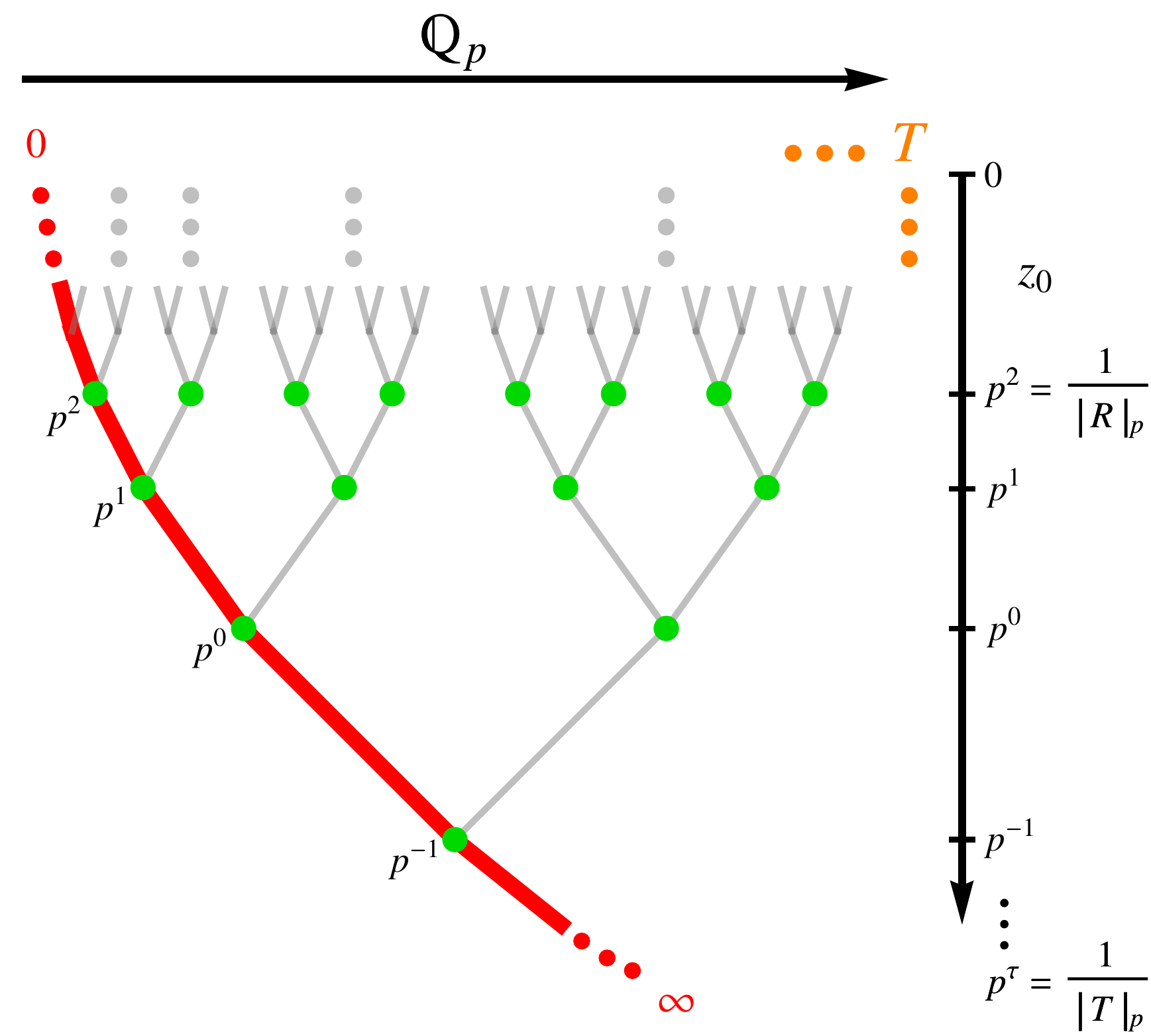}}
    \put(60,0){\includegraphics[clip,trim=0cm 3.2cm 0cm 0cm, width=4.5in]{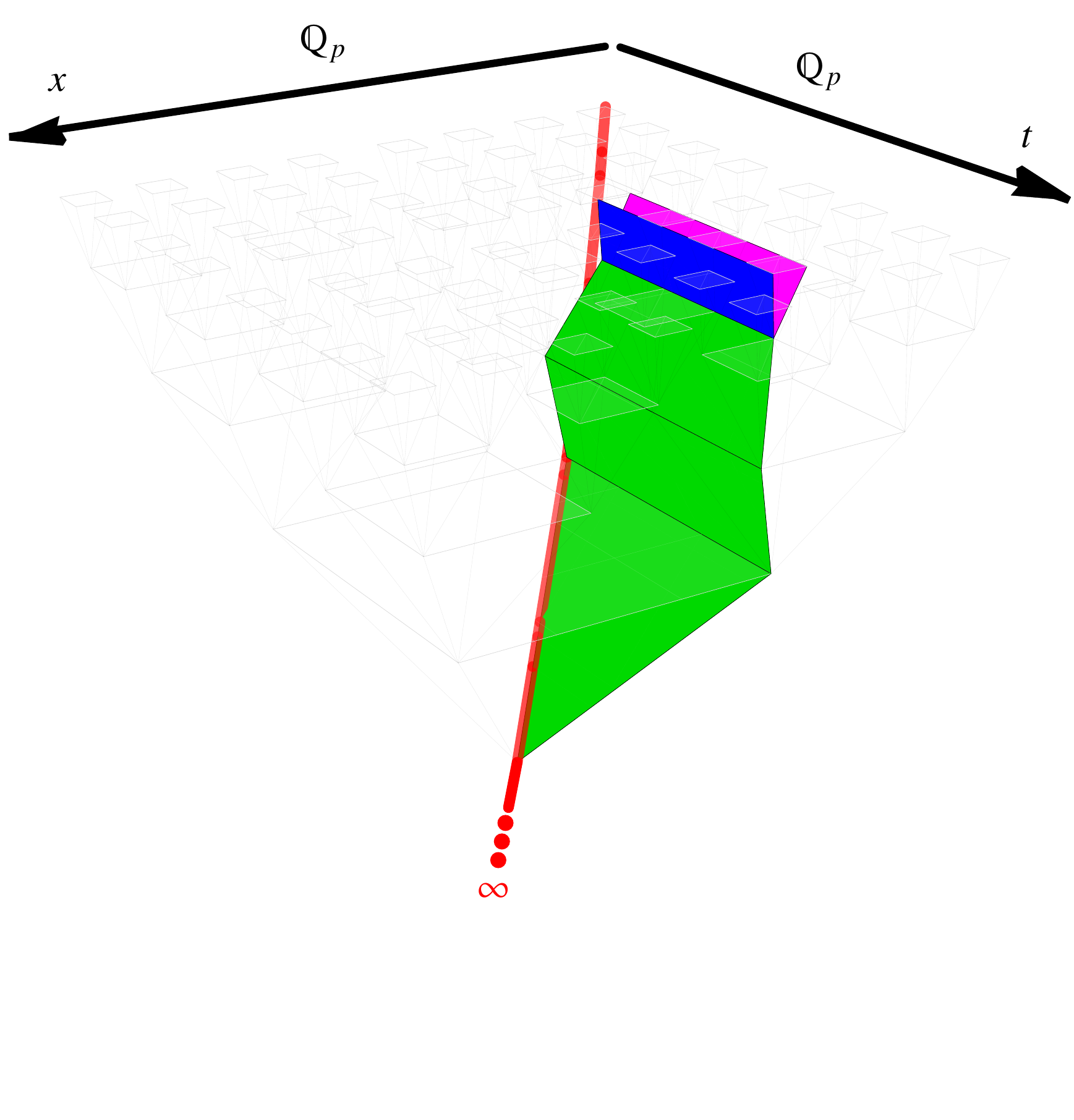}}
   \end{picture}
  \caption{(Color online.)  Two views of a Wilson loop in the Bruhat--Tits tree $T_q$ for $q=2^2$.  Top: Constant-$t$ and constant-$x$ slices of a Wilson loop with length $|T|_p = p^{-\tau}$ and $|R|_p = p^{-2}$.  Bottom: Segment of a Wilson loop in a perspective view.  The green region in both views indicate the vertices that are in common for the Wilson lines of the free quark and antiquark.  The purple region pertains to the quark, and the blue region pertains to the antiquark.}\label{figWilson}
 \end{figure}

The obvious $p$-adic analog of the Nambu--Goto action is the number of vertices on the string worldsheet.  More precisely, we want to count each vertex on $M_{q\bar{q}}(T)$ with multiplicity $1$, and at the same time count each vertex on $M_x(T) \sqcup M_{\tilde{x}}(T)$ with multiplicity $-1$.  As shown in figure~\ref{figWilson}, this counting is made easier by the observation that $M_{q\bar{q}}(T)$ covers precisely the points in $M_x(T) \cup M_{\tilde{x}}(T)$ down to a depth $z_0 = p^\rho$, so points above this depth (that is, points with $|z_0|_p < p^{-\rho}$) can be ignored.  Right at $z_0 = p^\rho$, where $M_{q\bar{q}}(T)$, $M_x(T)$, and $M_{\tilde{x}}(T)$ intersect, we should count points with net multiplicity $-1$ (which comes from $1$ for $M_{q\bar{q}}(T)$ plus $-1$ for each of $M_x(T)$ and $M_{\tilde{x}}(T)$).  Below this depth, continuing down to the infrared cutoff $z_0 = p^\tau$, we should count points with net multiplicity $-2$.  At a depth $m \geq \tau$ (meaning closer to the boundary than the infrared cutoff), the number of points on $M_x(T)$ (or $M_{\tilde{x}}(T)$) is $p^{m-\tau}$.  Thus the total count of points, including multiplicities as just described, is
 \eqn{PointsRegulated}{
  S_{\rm reg} = -p^{\rho-\tau} - 2\sum_{m=\tau}^{\rho-1} p^{m-\tau}
    = -\left| {T \over R} \right|_p {p+1 - 2 |R/T|_p \over p-1}  
    \approx -{\zeta_p(1)^2 \over \zeta_p(2)} \left| {T \over R} \right|_p \,,
 }
where in the last term we have dropped a term which is suppressed by a relative factor of $|R/T|_p \ll 1$.  The scaling of \eno{PointsRegulated} with $|T|_p$ and $|R|_p$ is as expected for a conformal theory, so that the potential $V(R) \propto 1/|R|_p$.  The coefficient in \eno{PointsRegulated} is negative, so that quarks and antiquarks in the dual theory attract, but it does not seem closely related to the prefactor in \eno{VArchimedean}.  Possibly a better understanding of more general Wilson loops could help shed light on this apparent mismatch.

\subsection{Future directions}
\label{Future}

There are a number of potentially interesting directions for further work.  To begin with, a more thorough analysis of the symmetries of $p$-adic field theories should be interesting.  We started with the simplest possible lattice action invariant under the isometries of $T_q$, and from it we derived correlators with some version of $p$-adic conformal symmetry.  We should ask, what exactly are the symmetries of these correlators?  Is the symmetry group simply ${\rm PGL}(2,\mathbb{Q}_q)$?  Or is there a larger symmetry algebra analogous to the Virasoro algebra?  How much of the structure of correlators is fixed by symmetry considerations?  For example, how much is the four-point function constrained by symmetry?  We should also ask whether we can proceed beyond scalar fields on the Bruhat--Tits trees and correspondingly scalar operators in the field theory.  If we start with more sophisticated lattice models on $T_q$, might we obtain correlators which can be expressed as products of multiplicative characters other than $|x|_p^s$?

Another interesting avenue to pursue is loop corrections.  All our calculations have relied on treating the bulk theory as classical, meaning that we focus on the specific field configuration which extremizes the action.  Adding in fluctuations perturbatively does not seem impossible, but the more interesting prospect is to pass to a fully statistical mechanical account of the bulk, where a ``temperature'' is dialed up from $0$, where our classical account is justified, to arbitrary finite values.  This may allow a closer connection with \cite{Harlow:2011az} as well as earlier works including \cite{Baxter:1982zz,Zinoviev:1990}.  Ideally, it may help us understand deeper connections with fluctuating fields in Archimedean AdS/CFT.  The real prize, of course, is to understand fluctuating geometry.  We suspect it is necessary to go beyond the Bruhat--Tits tree in order to properly formulate questions about dynamical geometry.  Perhaps the refinements of $T_q$ introduced in \eno{pAdSDef}--\eno{zzRequire} will be of help in this regard.

Yet another direction to explore is the full range of possible extensions of $\mathbb{Q}_p$.  What is the most natural notion of dimension in a boundary theory formulated on a (partially) ramified extension of $\mathbb{Q}_p$?  What sorts of correlators are sensitive to the particular extension we pick?  How is information about the extension encoded in the tree structure and in natural bulk actions? Moreover, following~\cite{Ghoshal:2006zh}, (totally) ramified extensions seem to represent a refinement of the Bruhat--Tits tree and may give more direct access to the Archimedean limit of $p$-adic results. Such a refinement seems to be of a different nature than the refinements of $T_q$ based on the geometry of chordal distance.  Could they however be related?

We hope to report on these questions in future work.

\clearpage

\section*{Acknowledgments}

S.S.G.~thanks P.~Landweber for useful discussions.  This work was supported in part by the Department of Energy under Grant No.~DE-FG02-91ER40671. The work of P.W.~was supported by the National Science Centre grant 2013/11/N/ST2/03812. The work of J.K.~was supported in part by a Princeton/DAAD Exchange Fellowship.

\bibliographystyle{ssg}
\bibliography{padic} 
\end{document}